\documentclass{article} 
\usepackage{iclr2026/iclr2026_conference,times}

\usepackage{amsmath,amsfonts,bm}

\def\eqref#1{equation~\ref{#1}}

\def\floor#1{\lfloor #1 \rfloor}
\def\1{\bm{1}}

\def\vd{{\bm{d}}}

\def\vs{{\bm{s}}}

\def\vu{{\bm{u}}}

\def\vw{{\bm{w}}}
\def\vx{{\bm{x}}}
\def\vy{{\bm{y}}}
\def\vz{{\bm{z}}}

\def\mI{{\bm{I}}}

\def\mK{{\bm{K}}}

\def\mZ{{\bm{Z}}}

\def\mSigma{{\bm{\Sigma}}}

\DeclareMathAlphabet{\mathsfit}{\encodingdefault}{\sfdefault}{m}{sl}
\SetMathAlphabet{\mathsfit}{bold}{\encodingdefault}{\sfdefault}{bx}{n}

\def\gN{{\mathcal{N}}}

\def\sC{{\mathbb{C}}}

\def\sR{{\mathbb{R}}}

\usepackage{url}

\usepackage{booktabs}
\usepackage{xspace}
\usepackage{graphicx}
\usepackage{wrapfig}
\usepackage[percent]{overpic}
\usepackage{algorithm}
\usepackage{algpseudocode}
\usepackage{caption}
\usepackage{subcaption}
\usepackage{enumitem}
\usepackage{tikz}
\usepackage{xcolor}
\usepackage[table]{xcolor}
\usepackage{tabularx}
\usepackage{multirow}
\usepackage{etoc}
\definecolor{deeppurple}{rgb}{0.1, 0.0, 0.45}
\definecolor{darkgreen}{rgb}{0, 0.7, 0}
\definecolor{darkblue}{rgb}{0, 0, 1}
\definecolor{myred}{HTML}{B12421}
\definecolor{myblue}{HTML}{15007E}
\definecolor{mypurple}{HTML}{651250}
\definecolor{purple}{HTML}{AD49E1}
\definecolor{teal}{HTML}{56DFCF}
\definecolor{pink}{HTML}{FF2DD1}

\long\def\Rebuttaltwo#1{\textcolor{black}{#1}}
\newcommand{\Rebuttalfour}[1]{{\color{black}#1}}

\newcommand{\Rebuttaleight}[1]{{\color{black}#1}}
\newcommand{\name}{\textsc{LatentFT}\xspace}
\newcommand{\myparagraph}[1]{\vspace{-4pt}\paragraph{#1}}

\DeclareMathOperator{\Enc}{Enc}
\DeclareMathOperator{\Dec}{Dec}
\DeclareMathOperator{\DFT}{DFT}
\DeclareMathOperator{\IDFT}{IDFT}
\DeclareMathOperator{\DiffusionForward}{DiffusionForward}

\algtext*{EndFor}
\algrenewcommand\algorithmicrequire{\textbf{Input:}}
\algrenewcommand\algorithmicensure{\textbf{Output:}}
\usepackage{hyperref}

\title{Latent Fourier Transform}

\author{Mason L. Wang \\
CSAIL \\
Massachusetts Institute of Technology \\
\texttt{ycda@csail.mit.edu} \\
\And
Cheng-Zhi Anna Huang \\
CSAIL \\
Massachusetts Institute of Technology \\
\texttt{huangcza@mit.edu} \\
}

\iclrfinalcopy 
\begin{document}

\maketitle

\etocdepthtag{main}

\begin{abstract}
    We introduce the Latent Fourier Transform (\name), a framework that provides novel frequency-domain controls for generative \Rebuttalfour{music} models. \name combines a diffusion autoencoder with a latent-space Fourier transform to separate musical patterns by timescale. By masking \Rebuttaltwo{latents in the frequency domain} during training, our method yields representations that can be manipulated coherently at inference. This allows us to generate musical variations and blends from reference examples while preserving characteristics at \Rebuttaltwo{desired} timescales, \Rebuttaltwo{which are specified as frequencies in the latent space.} \name parallels the role of the equalizer in \Rebuttalfour{music} production: while traditional equalizers operates on audible frequencies to shape timbre, \name operates on \Rebuttaltwo{latent-space} frequencies to shape musical structure. Experiments and listening tests show that \name improves condition adherence and quality compared to baselines. We also present a technique for hearing \Rebuttaltwo{frequencies in the latent space} in isolation, and show different musical attributes reside in different regions of the latent spectrum. Our results show how frequency-domain control in latent space provides an intuitive, continuous frequency axis for conditioning and blending, advancing us toward more interpretable and interactive generative \Rebuttalfour{music} models.
\end{abstract}
\section{Introduction}
\label{sec:intro}
%




Modern audio generation models often operate in a coarse-to-fine manner, generating progressively finer representations of the output signal in a conditional chain. In diffusion models \citep{kong2020diffwave, liu2023audioldm, huang2023noise2music}, higher noise levels provide coarser representations, while lower noise levels provide finer representations. In autoregressive models like AudioLM \citep{borsos2023audiolm} and MusicLM \citep{agostinelli2023musiclm}, an encoding stage represents the input signal as a hierarchy of coarse-to-fine tokens, and a generative model attempts to predict fine tokens from coarser ones. This is also the case for masked token models \citep{garcia2023vampnet}, discrete diffusion \citep{yang2023diffsound}, and next-scale prediction \citep{qiu2024efficient}.






Since the generative process involves conditioning on coarse representations, it is natural to generate new samples using the coarse representations of a reference example. This type of conditioning has been used for stroke-based image editing and image translation \citep{meng2021sdedit, choi2021ilvr}. 
However, conditioning on small- or mid-scale features is harder, since the representations used by the generative model rarely capture these features in isolation. For instance, discrete representations define fine tokens \emph{relative} to coarse ones via residual vector quantization (RVQ) \citep{zeghidour2021soundstream, kumar2023high}, preventing them from being interpreted independently. 



Conditioning on arbitrary timescales from a reference example would be useful in music, which contains slow-moving patterns (like chord progressions) and fast-moving patterns (like trills). Patterns occurring at different timescales may be desirable starting points for generating musical variations, but are difficult to specify precisely using text. Existing reference-based controls \citep{villa2025subtractive, garcia2025sketch2sound} target attributes like pitch, loudness, and instrumentation, which are distributed across multiple timescales. While these methods provide control over various semantic axes, none directly expose the `timescale' axis.

To address this, we explore the use of the Fourier transform,
which provides a decomposition of a signal into oscillations at different frequencies. High frequencies capture the most rapid variations in the signal (`small-scale' characteristics), while low frequencies capture slow variations in the signal (`large-scale' characteristics). This representation has two benefits:

\begin{itemize}[leftmargin=2.8em]
    \item First, frequency components are orthogonal, meaning that changing the signal's representation at one frequency does not affect the signal's representation at other frequencies. Thus, the Fourier transform provides an inductive bias for separating information across timescales.
    \item Second, the frequency axis provides an intuitive, continuous axis for specifying timescales precisely. The user can select for patterns based on the timescales \emph{in Hz} at which they occur, instead of relying on heuristic approaches for timescale specification.
\end{itemize}


Our approach merges the Fourier transform with deep representation learning: we use a diffusion autoencoder \citep{preechakul2022diffusion} to \emph{capture} musical patterns, and a latent-space Fourier transform to \emph{separate} them by scale. To achieve synergy between these two components, we propose a simple end-to-end training framework: an encoder transforms audio into a time series of latent vectors, which is randomly masked in the Fourier domain. Then, a decoder attempts to use this frequency-masked latent sequence to reconstruct the audio with a diffusion-based objective.

After training, we can encode user-selected music into a sequence of latent vectors. Then, we can apply a Fourier transform to this latent sequence, creating a \emph{latent spectrum}. \Rebuttaltwo{The latent spectrum maps different musical patterns to different frequencies in it, which we refer to as \emph{latent frequencies}. These latent frequencies correspond to the timescales at which the musical patterns occur.} The user can \emph{hear} different parts of the latent spectrum in isolation, or \emph{generate} variations while conditioning on patterns at \Rebuttaltwo{desired} timescales, \Rebuttaltwo{which are specified as latent frequencies}. Separation between timescales also allows us to \emph{blend} two musical examples together, retaining features at user-selected timescales from each. 
In short, we introduce novel frequency-based controls for generative models. 



To explain these controls and their effects, we draw parallels between our framework and the \emph{equalizer} (EQ), an essential tool in audio signal processing. The equalizer manipulates the \emph{audible spectrum}, or the frequencies in the audio \emph{waveform} within the limits of human hearing (20 -- 20,000 Hz). This shapes sonic characteristics like 
``warmth," ``brightness," ``clarity," and ``shine," which relate to different frequency ranges \citep[pp.~223--232]{izhaki2017mixing}. The equalizer is particularly crucial for \emph{mixing} multiple musical elements together coherently, by highlighting frequencies from each element and ensuring that elements do not ``clash'' over similar frequency ranges
\citep[pp.~14, 160--161]{owsinski2017mixing}. Since the equalizer operates on audio \emph{waveform} frequencies, it is unable to change musical or structural patterns (like notes or chords). These are more complex than waveform oscillations, and unfold on temporal scales below 20 Hz, where such oscillations are inaudible. Still, these structural patterns are also vital to combining multiple musical elements together in a coherent way. 

By operating on the \emph{latent spectrum} instead of the \emph{audible spectrum}, our framework provides a complement to the traditional equalizer that operates on musical patterns instead of sonic qualities. For instance, we can \emph{blend} sounds together in musically coherent ways, while preserving patterns from each sound at user-specified latent frequencies. This is akin to the way traditional EQs are used to \emph{mix} sounds together in musically pleasant ways, by choosing which audible frequencies of each sound to highlight. We dub our framework \name, and show several applications:



 \begin{enumerate}[leftmargin=2.8em]
     \item 
     \name can generate musically coherent variations of a given song, while preserving patterns at \Rebuttaltwo{desired timescales. These timescales are specified as  a mask over the latent frequency spectrum.} (Sec.~\ref{sec:cond}).
     \item 
     \name can blend two songs, preserving patterns from each at \Rebuttaltwo{desired timescales. These timescales are specified as masks over the latent frequency spectrum.}  (Sec.~\ref{sec:blend}).
     \item We can `zoom-in' on parts of the latent spectrum, allowing us to \emph{hear} musical patterns at \Rebuttaltwo{desired timescales, which are specified as latent frequencies} (Sec.~\ref{sec:iso}).
     \item We can interpret the latent spectrum of a song, and show where various musical characteristics like genre, tempo, and pitch reside on the \Rebuttaltwo{latent} spectrum (Sec.~\ref{sec:interpretability}).
\end{enumerate}
We demonstrate these applications through quantitative metrics (Table~\ref{tab:condandblend}), listening tests (Sec.~\ref{sec:listening}), and qualitative examples, which can be found on our \href{https://masonlwang.com/latentfouriertransform/}{website}\footnote{\url{https://masonlwang.com/latentfouriertransform/}}.

\section{Related Work}
\label{sec:related}
\myparagraph{Audio Generation.}
Recent years have witnessed a great expansion in audio-domain generative models, which operate in a continuous domain or by generating discrete tokens.
Diffusion models \citep{sohl2015deep, ho2020denoising, song2020score} generate samples by iteratively denoising pure Gaussian noise.
Other approaches to audio generation rely on discrete audio codec tokens \citep{zeghidour2021soundstream, kumar2023high}, which compress audio into a multi-layer sequence of tokens, with successive layers capturing increasingly fine details. Token generation can proceed in an autoregressive \citep{borsos2023audiolm, copet2023simple, agostinelli2023musiclm} or non-autoregressive \citep{garcia2023vampnet, borsos2023soundstorm} manner, but in both cases, coarse tokens typically condition the generation of finer ones. We propose Fourier-based representations that 
let us condition on features at arbitrary scales. We compare our method to conditioning on intermediate or fine tokens in our \hyperref[baseline:vampnet]{Masked Token Model baseline}.


\myparagraph{Controls for Audio \Rebuttalfour{and Music} Generation.} Current audio generation methods \Rebuttalfour{offer} global controls like text \citep{Forsgren_Martiros_2022, huang2023noise2music, liu2023audioldm, copet2023simple, agostinelli2023musiclm, chen2024musicldm, schneider2024mousai, evans2025stable}, or time-varying controls based on musical attributes like pitch and loudness curves \citep{wu2024music, garcia2025sketch2sound} or stems \citep{parker2024stemgen, villa2025subtractive}. Different time-varying signals allow for control along different semantic axes, but not along the `timescale' axis. These works mostly condition on the \emph{entire} control signal, not selected frequency components. The exception is Sketch2Sound \citep{garcia2025sketch2sound}, which optionally smooths the pitch or loudness-based control signal using median filtering. Still, this type of filtering is heuristic, applies only to preserving large-scale features, and operates on hand-extracted features instead of latent ones. Guidance \citep{levy2023controllable} and initial noise optimization \citep{novack2024ditto} have also been used to control music generation using differentiable objectives. We use guidance for our tasks in our \hyperref[baseline:guidance]{Guidance baseline}. 

\myparagraph{Image Editing Frameworks.} 
The coarse-to-fine paradigm lends itself to image editing frameworks that generate variations of input examples based on their low-frequency features. SDEdit \citep{meng2021sdedit} enables stroke-based image generation and editing by adding 
white noise 
to a given reference (which acts like a heuristic low-pass filter), and running the denoising process. 
Similarly, Iterative Latent Variable Refinement (ILVR) \citep{choi2021ilvr} can generate variations of images while preserving large-scale structure. During the denoising process, ILVR continually replaces the low-frequency components of the noisy sample with the low-frequency components of a (noised) reference, enabling image translation and stroke-based editing. ILVR does not condition on high-frequency or mid-frequency components, but we attempt this in our \hyperref[baseline:ilvr]{ILVR baseline}.  




\myparagraph{Fourier-Based Deep Learning.}
While we apply the Fourier transform to latent vectors, many works use frequency-domain representations of the input or output space. These include works in vision \citep{lee2018single, yang2020fda, atzmon2024edify} and audio \citep{san2023discrete, moliner2024diffusion}.
Similar to our method (Sec.~\ref{sec:method}),
\cite{zheng2024mfae} propose a frequency-masked autoencoder that extends the masked image modeling paradigm \citep{he2022masked, xie2022simmim} to the frequency domain. \Rebuttaleight{AudioMAE \citep{huang2022masked} applies masked image modeling to audio spectrograms, randomly masking time-frequency bins in the audio spectrogram domain.} However, our method masks \emph{latent\Rebuttaltwo{-space}} frequencies.

Other works \emph{do} apply the Fourier transform to hidden states, but do so as part of black-box architectural units, and focus on downstream tasks instead of directly using the latent spectra. This use of the Fourier transform has been shown to improve learning in language \citep{lee2021fnet, he2023fourier} and vision \citep{rao2021global, chi2020fast, guibas2021adaptive, lin2023deep}.

Finally, some works apply the Fourier transform \emph{post-hoc} to latent states of pretrained models, 
\emph{choosing} and \emph{interpreting} latent\Rebuttaltwo{-space} frequencies. PRISM \citep{tamkin2020language} shows that different frequency bands of language model embedding sequences are useful for different downstream tasks. In vision, \cite{khan2017scene} shows that the spectra of intermediate activations in a pretrained CNN can be used to categorize scenes. These works focus on \emph{analysis}, while we focus on \emph{synthesis}: we can isolate frequencies in the latent representation, but also \emph{invert} them and observe their realizations in the input domain. Applying frequency-domain manipulations \emph{post-hoc} to pretrained representations fails to \emph{synthesize} coherent audio, which we show in the \hyperref[baseline:dac]{DAC} \Rebuttaleight{and \hyperref[baseline:rave]{RAVE}} baselines and our ablations (Appendix ~\ref{app:ablations}). This shortcoming motivates our frequency-masking strategy during training, which deliberately encourages our latents to be manipulable in the frequency domain.

\myparagraph{Blending.} \name can blend two examples together while \Rebuttaltwo{choosing} timescales from each \Rebuttaltwo{(by selecting latent frequencies from each example)}.
This is like style transfer in images \citep{ashikhmin2003fast, gatys2016image,johnson2016perceptual,huang2017arbitrary,deng2022stytr2, efros2023image}, which merges ``content" from one image with the ``style" from another. Applying these methods to music is challenging due to the \emph{multiscale} nature of musical style, as style can refer to ``high-level compositional features" or ``low-level acoustic features" \citep{dai2018music}. We ameliorate this ambiguity by introducing \emph{frequency-based} controls, which provide a continuous axis for specifying which timescales we want from each input. In contrast, existing works in musical style transfer focus on specific aspects of music like timbre \citep{huang2018timbretron, li2024music, wang2024training}, musical arrangement \citep{cifka2020groove2groove}, or composition \citep{se2016musical}. Traditional techniques are also used to blend sounds, as done in the \hyperref[baseline:cross]{Cross Synthesis baseline}~\citep{smith2011spectral}.

\section{Method}
\label{sec:method}
\begin{wrapfigure}{r}{0.22\linewidth}
    \vspace{-40pt}
    \centering
    \begin{overpic}[width=\linewidth, trim=0 7 7 0, clip]{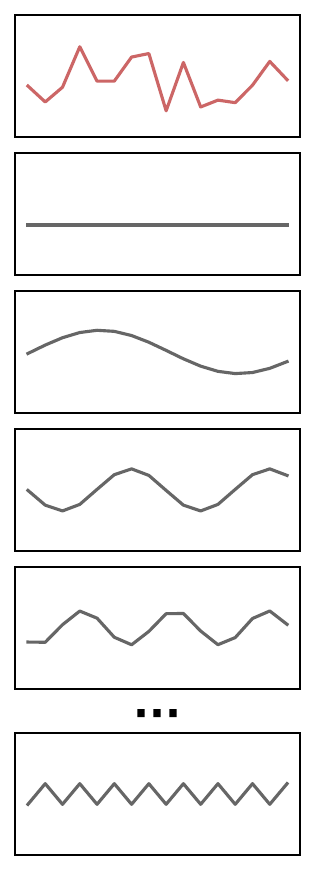}
    \put(2.5,95.5){\scriptsize$\vx$}
    \put(2.5,79.25){\scriptsize$k=0$}
    \put(2.5,63){\scriptsize$k=1$}
    \put(2.5,46.75){\scriptsize$k=2$}
    \put(2.5,30.5){\scriptsize$k=3$}
    \put(2.5,11.3){\scriptsize$k=8$}
    \end{overpic}
    \caption{$\vx \in \sR^{16}$ decomposed via Eq~\ref{eq:realidft}.}
    \vspace{-30pt}
\label{fig:sinusoids}
\end{wrapfigure}
\subsection{Background}
\myparagraph{Discrete Fourier Transform.}
The discrete Fourier transform\footnote{We present a simplified notation for the DFT for clarity and brevity. Note that $\vw_k$ is typically presented as the complex conjugate of what we have here.} (DFT) correlates an input signal $\vx \in \sC^N$ with $N$ complex sinusoidal signals, giving its spectral representation $\bm{X} \in \sC^N$. The $k$th DFT coefficient is given by:
\begin{equation}
    \bm{X}[k] = \vx \cdot \bm{w}_{k}, \tag{DFT}
    \label{eq:dft}
\end{equation}
where $(\cdot)$ denotes the complex dot product, and $\vw_k[n] = e^{j(2\pi k/ N)n}$ denotes the $k$th complex sinusoid. The complex sinusoids ${\vw_1, ... , \vw_N}$ form an orthogonal basis for $\sC^N$, allowing the DFT to be inverted:
\begin{equation}
    \vx = \frac{1}{N} \sum_{k=0}^{N-1} \bm{X}[k] \bm{w}_{k}  \tag{IDFT}
    \label{eq:idft}
\end{equation}
The inverse DFT is also called the ``synthesis" equation, since it expresses $\vx$ as a weighted sum of complex sinusoids. To provide more concrete intuition, if $\vx$ is real-valued, we can express $\vx$ as the sum of \emph{real} sinusoids with various frequencies $\frac{k}{N}$, amplitudes $A_k$,  and phase shifts $\phi_k$:
\begin{equation}
    \vx[n] =  \sum_{k=0}^{\floor{N/2}} A_k \cos \left( 2 \pi \frac{k}{N}  n + \phi_k \right)
    \label{eq:realidft}
\end{equation}
Where $A_k$ and $\phi_k$ are both derived from the coefficient $\bm{X}[k]$, as shown in Appendix~\ref{app:dft}. In words, the DFT can decompose a \emph{real} signal into a sum of \emph{real} sinusoids of different frequencies, all of which are mutually orthogonal. We show this decomposition for an example signal in Fig.~\ref{fig:sinusoids}.

\myparagraph{Diffusion Autoencoders.}
The diffusion autoencoder was proposed by \citep{preechakul2022diffusion} to harness the power of diffusion models for representation learning. \Rebuttaltwo{During training}, an encoder maps an image $\vx_0$ into a non-spatial semantic vector $\vz_{\text{sem}}$. Then, a diffusion model (which acts as the decoder) \Rebuttaltwo{tries to reconstruct} $\vx_0$ from $\bm{z}_{\text{sem}}$ and a noisy version of the image $\vx_{\tau}$. \Rebuttaltwo{Diffusion autoencoders are typically trained with a MSE loss that determines how well $\vx_{\tau}$ is denoised, (or equivalently, how well $\vx_0$ is reconstructed). During inference, $\vz_{\text{sem}}$ can be used to condition a generative diffusion process and produce an image.
}

\Rebuttaltwo{We have three motivations for using a diffusion autoencoder. First, the decoder harnesses the generative power of a diffusion model, allowing it to generate high-quality music even when information has been removed (masked) from the latent conditioning vector. Second, since the generative process is random, one can generate multiple variations for the same input condition. Third, diffusion autoencoders have been shown to yield latent representations $\vz_{\text{sem}}$ that are semantically meaningful and linear, supporting interpolation between images and attribute manipulation. In fact}, recent work shows the applicability of diffusion autoencoders to
music representation learning \citep{pasini2024music2latent, bindi2024unsupervised}.

\subsection{Method Overview}
\begin{figure}[t]
    \centering
    \includegraphics[width=\linewidth, trim=0 0 0 0, clip]{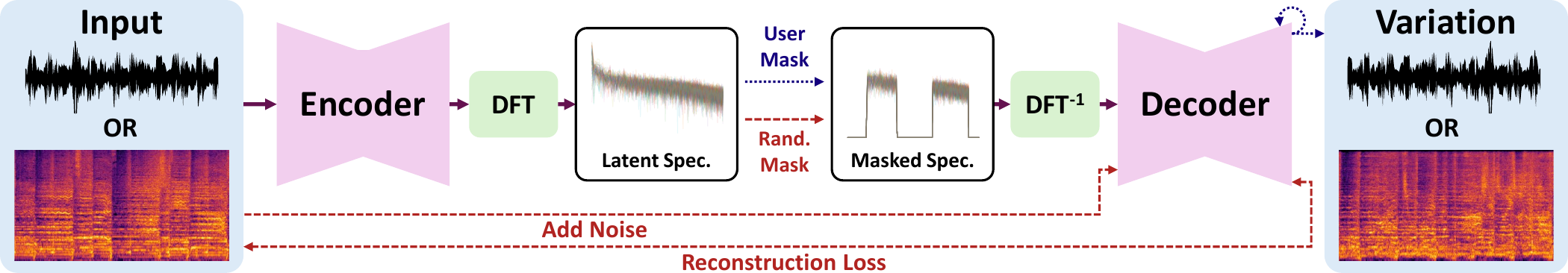}
    \caption{\textbf{Latent Fourier Transform (\name).} We encode audio \Rebuttaltwo{(which may be represented as a waveform or spectrogram)} into a series of latent vectors and compute a latent spectrum. During training \textcolor{myred}{(red)}, this spectrum is masked randomly and used to reconstruct the input. During inference \textcolor{myblue}{(blue)}, the user specifies a spectral mask, which selects features from the input at specific \Rebuttaltwo{latent frequencies} and conditions a generative process.}
    \vspace{-10pt}
    \label{fig:method}
\end{figure}
Our goal is two-fold. First, we want to map an audio waveform \Rebuttaltwo{or spectrogram} $\vx_0$ into a \Rebuttaltwo{time} series of latent vectors, whose spectrum encodes semantic patterns. \Rebuttaltwo{We refer to the DFT spectrum of this latent time series as the \emph{latent spectrum}. It is important to distinguish the latent spectrum from the audible spectrum: The audible spectrum refers to the DFT spectrum of the audio waveform, and captures variations in the waveform occurring at different frequencies. In contrast, the latent spectrum captures variations in the latent time series occurring at different frequencies, which we correspond to musical patterns occurring at different timescales.
}
Second, we should be able isolate features at selected \Rebuttaltwo{latent frequencies} and use them to \emph{generate variations}, blend them with other audio clips, or hear them in isolation. These goals motivate an end-to-end \emph{encoder-decoder} architecture that \emph{encodes} music into latent spectra, and \emph{decodes} latent spectra into music. 
We apply a latent Fourier transform and frequency-masking \emph{during training}, shown in Alg.~\ref{alg:training} and Fig.~\ref{fig:method}.

\subsection{Encoding the Latent Spectrum}

\myparagraph{Encoder.} An encoder maps input music $\vx_0 \in \sR^{C \times T}$ to a time series of latent vectors $\vz \in \sR^{C' \times T'} $:
\begin{equation}
    \vz = \Enc_{\phi}(\vx_0)
\end{equation}
\begin{wrapfigure}{r}{0.47\linewidth}
\vspace{-20pt}
\resizebox{6.57cm}{!}{
\small
\begin{minipage}{\linewidth}
\begin{algorithm}[H]
\small
\captionsetup{font=small}
\caption{\small{Training.}}
\begin{algorithmic}[1]
\Require Audio Waveform or Spectrogram $\vx_0$
\State $\vz \gets \Enc_{\phi}(\vx_0)$
\State $\bm{Z} \gets \DFT(\vz)$
\State $\eta \sim \mathcal{N}(0, 1)$ \Comment {Sample threshold}
\State $\vs \sim \mathcal{N}(\bm{0}, \mSigma)$ \Comment {Sample frequency bin scores}
\State $\bm{M} \gets \1_{\vs > \eta}$ \Comment {Get Mask}
\State $\bm{Z}^{\text{masked}} \gets \bm{Z} \odot \bm{M}$
\State $\vz^{\text{masked}} \gets \IDFT\left(\bm{Z}^{\text{masked}}\right)$
\State Sample noise level $\tau \sim p(\tau)$
\State $\vx_{\tau} \gets \DiffusionForward \left( \vx_0 , \tau \right)$ \Comment{Add noise}
\State $\hat{\vx}_{0} \gets \Dec_{\theta}\left(\vz^{\text{masked}}, \vx_\tau, \tau \right)$  \Comment{Reconstruct $\vx_0$}
\State $\ell \gets \mathcal{L}\left(\hat{\vx}_{0}, \vx_0 \right) $
\State Update parameters $\phi, \theta$ using $\nabla_{\phi, \theta} \ell$
\end{algorithmic}
\label{alg:training}
\end{algorithm} 
\end{minipage}
}
\vspace{-10pt}
\end{wrapfigure}
Here, $C$ and $C'$ are the number of input and latent channels, while $T$ and $T'$ are the number of input and latent timesteps. Although $T$ and $T'$ do not have to be equal, $\vz$ must have a linear temporal axis in order to produce a latent spectrum. This favors convolutional architectures or networks like the U-Net \citep{ronneberger2015u}, whose skip connections promote input-output alignment. \Rebuttaltwo{We define $f_r$ as the latent frame rate in Hz, or the number of latent vectors (frames) needed to represent one second of audio.}


\paragraph{Latent Fourier Transform.} \Rebuttaltwo{The \emph{latent spectrum}\footnote{The DFT is different from the Short-time Fourier Transform, which yields a \emph{spectrogram}, not a \emph{spectrum}.} refers to the DFT of the latent timeseries $\vz$, applied to} each channel in the latent timeseries:
\begin{equation}
    \bm{Z} = \DFT(\vz), \quad \bm{Z} \in \sC^{C' \times K}
\end{equation}
Applying the DFT along the time axis of our latent sequence represents each latent channel as a sum of $\Rebuttaltwo{K} = \floor{T'/2} + 1$ sinusoids (see Eq.~\ref{eq:realidft}). \Rebuttaltwo{The sinusoids have $K$ different linearly-spaced frequencies, which} capture variations in each latent channel at different \Rebuttaltwo{temporal rates}. The $k$th sinusoid completes $k$ cycles in $T'$ latent timeframes (see Fig.~\ref{fig:sinusoids}). \Rebuttaltwo{For instance, the $0$th sinusoid is constant, the $1$st sinusoid has a period of $T'$ latent frames, and the $2$nd sinusoid has a period of $T'/2$ latent frames. The sinusoids} are also orthogonal from one another, creating an inductive bias for separating information across timescales. 

\Rebuttaltwo{Specifically, $\DFT(\vz)$ stores $K$ complex coefficients indicating the amplitude and phase of each sinusoid along a length-$K$ frequency axis. We refer to the frequency-axis of $\DFT(\vz)$ as the \emph{latent frequency axis}, and we call points along this axis \emph{latent frequencies.} 
Like audible frequencies, latent frequencies are described in Hz. However, 1 Hz on the latent spectrum corresponds to oscillations in the \emph{latent sequence} occurring at 1 cycle per second, instead of oscillations in the audio waveform. The $k$th sinusoid has a period of $T'/k$ latent frames or $T'/(k f_r)$ seconds, and thus a latent frequency of $f_k =  kf_r/T'$ Hz.}

\myparagraph{Increasing Spectral Granularity.} In practice, we zero-pad $\vz$ at its end, expanding its temporal length by a factor of $L$. This increases the number of frequency bins by a factor of $\approx L$, allowing for more spectral granularity via \emph{spectral interpolation} \citep{smith2007mathematics}. This is especially useful for capturing very low-frequency patterns (below 1 cycle per $T'$ timeframes).
We let $F = \floor{LT'/2} + 1$ be the number of spectral bins (sinusoids) \Rebuttaltwo{after zero padding $\vz$}.


\subsection{Frequency Masking}
\label{subsec:masking}
At inference, we want to select specific frequencies from the latent spectrum to generate variations from them or hear them isolation (`zoom-in' on them). This is accomplished by applying a latent spectral mask $\bm{M} \in \{0,1\}^{F}$, taking $\bm{Z}^{\text{masked}} = \bm{Z} \odot \bm{M}$.
During inference, this mask is chosen by the user. During training, this mask is \emph{randomized}: First, we sample a random scalar threshold $\eta \sim \mathcal{N}(0, 1)$, which helps decide the proportion of bins to be masked. Second, we sample $\vs \sim \mathcal{N}(\bm{0}_F, \mSigma)$, where $\vs \in \sR^F$ assigns scores to each frequency bin. Third, we set the mask to keep bins whose score is greater than the threshold, setting
$\bm{M} = \1_{\vs > \eta}$.
\myparagraph{Random Threshold. } Using a random threshold ensures a uniform distribution over the number of masked bins. In contrast, independently masking each frequency bin with probability $p$ corresponds to setting a fixed threshold and $\mSigma = \mI$. This results in a binomial distribution over the number of masked bins, which does not reflect the inference-time distribution of user-specified masks.

\myparagraph{Correlating Bins.} 
Instead of masking each frequency bin independently, we create a ``soft grouping" between nearby frequency bins by correlating their scores. This is done by multiplying uncorrelated scores $\vu \sim \mathcal{N}(\bm{0}_F, \mI$) with a radial basis function matrix $\mK$:
\begin{equation}
    \mK_{i,j} = c_i \exp \left( -\frac{|a_i - a_j|^p}{2\sigma^p} \right), \quad \mK \in \sR^{F \times F},
    \label{eq:correlation}
\end{equation}
where $a_i = \log(f_i + \epsilon)$ is the frequency of bin $i$ mapped to a logarithmic axis, $p, \sigma$, and $\epsilon$ are hyperparameters, and $c_i$ normalizes each row of $\mK$ to have unit $\ell_2$ norm. Multiplying $\vs = \mK \vu$ results in correlated scores between frequency bins, where the amount of correlation between two frequency bins is determined by their distance on a logarithmic axis. 
The covariance matrix of $\vs$ is $\mSigma = \mK \mK^T$. 
Ablations (Appendix~\ref{app:ablations}) show that correlating bin scores is key to our method's performance . Intuitively, masking frequency bins independently forms speckled masks where masked bins are often adjacent to unmasked ones. The unmasked bins provide strong local cues about nearby masked ones, 
reducing the model's ability to fill in 
contiguous regions of the latent spectrum during inference. In contrast, correlated bin scores form masks with larger contiguous regions, which combats the effect of spectral leakage and better reflects inference-time, user-specified masks.

Logarithmically scaling the frequency-axis is also key to performance (Appendix~\ref{app:ablations}). This is common in audio, exemplified by the Mel scale \citep{stevens1937scale}, Constant Q-Transform \citep{brown1991calculation}, and others.
More generally, structured signals from images \citep{san2023discrete} to coastlines and mountains \citep{bak1987self} have spectra that follow a $1/f^{\alpha}$ curve. Segmenting such spectra into groups of equal width along a log-frequency axis yields groups of roughly equal energy. This motivates our logarithmic scaling, where higher frequencies are more likely to form larger groups. Lastly, normalizing the rows of $\mK$ 
ensures equal marginal variance between every bin score $\vs_k$, so that all bins have the same marginal probability of being masked for any given threshold.

\subsection{Decoding the Latent Spectrum}
We transform $\mZ^{\text{masked}}$ back into the time domain by applying the inverse DFT, obtaining a frequency-masked latent sequence $\vz^{\text{masked}} = \IDFT(\bm{Z}^{\text{masked}}).
$
The decoder then uses $\vz^{\text{masked}}$ to reconstruct the input $\vx_0$ from a noisy version of it (training), or to condition a diffusion process (inference). 

\myparagraph{Training.}
During training, we obtain a noisy version $\vx_{\tau}$ of the input $\vx_0$ through a forward diffusion process. This process samples a diffusion time $\tau \sim p(\tau)$ from a predetermined distribution and adds a $\tau$-dependent amount of noise to $\vx_{0}$. We supply $\vz^{\text{masked}}$ and $\vx_{\tau}$ to the decoder, which gives an estimate of the clean input $\hat{\vx}_{0}$: 
\begin{equation}
    \hat{\vx}_{0} \gets \Dec_{\theta}\left(\vz^{\text{masked}}, \vx_\tau, \tau \right)
    \label{eq:decoding}
\end{equation}
Then, we compute a reconstruction loss $\ell = \mathcal{L}\left(\hat{\vx}_{0}, \vx_0 \right)$, which is used to update the parameters $\phi, \theta$ of both the encoder and decoder. This procedure effectively trains a diffusion model, which can generate new outputs conditioned on $\vz^{\text{masked}}$. While we do not require a particular diffusion framework, in practice we follow the ODE formulation in \cite{karras2022elucidating}. This framework preconditions the model inputs and outputs, uses approximately linear diffusion trajectories, and applies a second-order correction at each sampling step (omitted in Algs.~\ref{alg:conditional} and ~\ref{alg:blending} for clarity).


\begin{wrapfigure}{r}{0.43\linewidth}
\vspace{-22pt}
\resizebox{5.9cm}{!}{
\small
\begin{minipage}{\linewidth}
\begin{algorithm}[H]
\small
\captionsetup{font=small}
\begin{algorithmic}[1]
    \Require $\vz^{\text{masked}}, \{
    \tau_i \}_{i=0}^N
    $ decreasing
    \vspace{2pt}
    \State $\vx \sim \gN(\bm{0}, \sigma_{\text{max}}^2)$ 
    \For{$i \in \{0, ...,  N - 1\} $}
        \State $\hat{\vx}_0 \gets \Dec_{\theta}\left(\vz^{\text{masked}}, \vx, \tau_i \right)$
        \State $\vd \gets \left(\vx - \hat{\vx}_0\right)/\sigma_i$ \Comment{{\scriptsize Deriv. of Noise Traj.}}
        \State $\vx \gets \vx + \left(\tau_{i+1} - \tau_{i} \right) \vd$ 
    \EndFor
    \State \Return $\vx$
\end{algorithmic}
\caption{\small{Conditional Generation}}
\label{alg:conditional}
\end{algorithm} 
\vspace{-15pt}
\begin{algorithm}[H]
\small
\captionsetup{font=small}
\begin{algorithmic}[1]
    \Require $\vz_1^{\text{masked}}, \vz_2^{\text{masked}}, \{
    \tau_i \}_{i=0}^N$, weights $\alpha, \beta$ 
    \vspace{2pt}
    \State $\vx \sim \gN(\bm{0}, \sigma_{\text{max}}^2)$ 
    \For{$i \in \{0, \ldots,  N - 1\} $}
        \State $\hat{\vx}_0^{(1)} \gets \Dec_{\theta}\left(\vz^{\text{masked}}_1, \vx, \tau_i \right)$
        \State $\hat{\vx}_0^{(2)} \gets \Dec_{\theta}\left(\vz^{\text{masked}}_2, \vx, \tau_i \right)$
        \State $\vd_1 \gets (\vx - \hat{\vx}_0^{(1)})/\sigma_i$
        \State $\vd_2 \gets (\vx - \hat{\vx}_0^{(2)})/\sigma_i$ 
        \State $\vd \gets \alpha \vd_1 + \beta \vd_2$ 
        \State $\vx \gets \vx + \left(\tau_{i+1} - \tau_{i} \right) \vd$ 
    \EndFor
    \State \Return $\vx$
\end{algorithmic}
\caption{\small{Blending}}
\label{alg:blending}
\end{algorithm} 
\end{minipage}
}
\vspace{-20pt}
\end{wrapfigure}

\myparagraph{Conditional Generation. } Our conditional generation task attempts to generate a variation of a reference song $\vy$ that preserves characteristics at user-specified \Rebuttaltwo{latent frequencies}. The reference $\vy$ is encoded and masked in the latent frequency domain to obtain $\vz^{\text{masked}}$. The mask is user-specified, and typically selects low frequencies, high frequencies, or a band of intermediate frequencies. We use $\vz^{\text{masked}}$ to condition a reverse diffusion process, which iteratively denoises pure Gaussian noise 
to yield a new variation.

\myparagraph{Blending. } Our blending task attempts to combine two musical references $\vy_1, \vy_2$ into a new song that preserves characteristics from each at user-specified \Rebuttaltwo{latent frequencies}. Like before, $\vz_1,\vz_2$ are obtained and masked in the latent frequency domain to get conditions $\vz_1^{\text{masked}}$, and $\vz_2^{\text{masked}}$. Here, the user specifies \emph{two} masks specifying which latent frequencies to retain from \emph{each} input. We obtain our blend by simulating the reverse diffusion process, at each step interpolating the derivatives induced by each condition (Alg.~\ref{alg:blending}).








\section{Experiments}
\label{sec:experiments}

\subsection{Experimental Setup}
\myparagraph{Datasets and Training.}
We train \Rebuttalfour{three} versions of \name with \Rebuttalfour{three} different encoders. \Rebuttalfour{We use UNet and MLP encoders with a mel-spectrogram frontend, as well as a raw audio encoder that utilizes a Descript Audio Codec (DAC) \citep{kumar2023high} frontend. These encoders are described further and compared in Appendix~\ref{app:encoders}}. Each model generates mel-spectrograms, which are inverted using the BigVGAN neural vocoder \citep{lee2022bigvgan}. We train our model on MTG-Jamendo \citep{bogdanov2019mtg}, a large-scale collection of over 55,000 songs spanning diverse musical genres (described more in Appendix~\ref{app:datasets}) segmented into 5.9-second musical clips. \Rebuttaltwo{Hyperparameters for the decoders and training are in Appendix~\ref{app:decoders} and~\ref{app:training}, respectively.} 
\paragraph{Baselines.} We compare \name to various traditional and learned methods of generating or representing audio. First, we try several \emph{generation} baselines, adapting them to our task:
\begin{itemize}[leftmargin=2.4em]
    \item \textbf{Masked Token Model} \citep{garcia2023vampnet}.
    \label{baseline:vampnet}
    We use the Vampnet masked token model, which generates discrete acoustic tokens from coarse-to-fine. Vampnet is trained to predict all acoustic tokens given a random subset of them, and supports supplying arbitrary token masks during inference. For conditional generation, we select different contiguous subsets of RVQ layers to condition on, and for the blending task, we select a different layer to take from each reference.
    \item \textbf{Guidance}     \citep{levy2023controllable}.
    \label{baseline:guidance} We generate mel-spectrograms with an unconditional diffusion model. At each denoising step, we compute the DFT along the time axis of the reference spectrogram(s) and the current reconstruction $\hat{\vx}_0$. We compute the loss between these DFT spectra \emph{within} the selected frequency bins, using it to update the intermediate output.
    \item \textbf{ILVR} \citep{choi2021ilvr}. 
    \label{baseline:ilvr}
    We generate mel-spectrograms from an unconditional diffusion model. At each denoising step, we compute DFT spectra of the intermediate output and the reference(s) set to the current noise level. We replace selected DFT frequencies of the intermediate output with the corresponding DFT frequencies of the noisy reference(s).
    \item \textbf{Cross Synthesis} \citep{smith2011spectral}.
    \label{baseline:cross}
    Cross synthesis blends two sounds by replacing the spectral envelope of one sound with that of the other. We follow the implementation in \cite{smith2011spectral}.
\end{itemize}
In the Guidance and ILVR baselines, note that we use the spectrum of the mel-spectrogram to steer the diffusion process instead of the latent spectrum.
We also attempt \emph{post-hoc} frequency-domain filtering of existing \emph{representations} of audio for our tasks, similar to \cite{tamkin2020language}:
\begin{itemize}[leftmargin=2.4em]
    \item \textbf{DAC} \citep{kumar2023high}. 
    \label{baseline:dac}
    We encode our reference(s) using Descript Audio Codec, a popular deep neural audio codec. We frequency-mask the latent states post-quantization, and feed the filtered latent sequence to the decoder to produce audio.


    \Rebuttaleight{
    \item \textbf{RAVE} \citep{caillon2021ravevariationalautoencoderfast} offers another latent representation of the audio signal, which is often manipulated in the latent space and used to generate audio \citep{nabi2024embodied, zheng2024mapping}. Similar to DAC, we frequency-mask the latent states obtained from the RAVE encoder, then provide them to the decoder.
    \label{baseline:rave}
    }

    \item \textbf{Spectrogram.}
    \label{baseline:spectrogram}
    We filter the input mel-spectrogram representation(s) directly, \Rebuttalfour{by computing the DFT of the mel-spectrogram(s) along the time axis, then masking the DFT(s).} We convert the filtered mel-spectrograms to audio with BigVGAN \citep{lee2022bigvgan}.
\end{itemize}
In each case, we blend by taking selected frequency components from two latent representations derived from two inputs, by adding the two frequency-masked latents together before decoding.

\subsection{Conditional Generation}
\label{sec:cond}
We show that \name can generate variations of a given song while preserving patterns at user-specified timescales. We take 1024 random 5.9-second clips from the MTG-Jamendo test set, ensuring each clip originates from a unique song (results on more datasets in Appendix~\ref{app:more_datasets}). We then generate variations of each clip, conditioning on 14 different latent frequency bands of varying widths and locations (see Appendix~\ref{app:condandblend}). 
Good variations should \emph{adhere} to the condition, preserving input characteristics at the specified timescales, and have musically coherent, high-\emph{quality} audio.
\myparagraph{Metrics.} 
To measure adherence, we extract time-series descriptor signals (e.g., loudness curves) from both the input and generated audio. We bandpass these descriptor signals to the selected frequency band, and measure their similarity or error with standard metrics.
We select four perceptually relevant time series descriptors. First, we extract \emph{loudness} curves following \cite{morrison2024fine}, and quantify their similarity using their correlation coefficient \citep{kosta2016mapping}. Second, we quantify \emph{rhythmic} preservation by computing onset strength envelopes \citep{bock2013maximum} and measuring their beat-spectral cosine similarity \citep{foote2002audio}. Third, to quantify \emph{timbral} preservation, we extract Mel-Frequency Cepstral Coefficients and compute Mel-Cepstral Distortion \citep{kominek2008synthesizer}. Fourth, we measure \emph{harmonic} characteristics (relating to chords and music notes) by computing tonal centroid features, and quantify error using Tonnetz distance \citep{milne2016empirically}.
We measure audio quality by computing the Frechet Audio Distance \citep{kilgour2018fr} between the set of generated music and the MTG-Jamendo validation set.


\begin{table*}
\centering
\resizebox{\linewidth}{!}{%
\begin{tabular}{l cccc c  cccc c}
    \toprule
    %
    & \multicolumn{5}{c}{\textbf{Conditional Generation}}
    & \multicolumn{5}{c}{\textbf{Blending}} \\
    \cmidrule(lr){2-6}
    \cmidrule(lr){7-11}

    & \multicolumn{4}{c}{Adherence}
    & \multicolumn{1}{c}{Quality}
    & \multicolumn{4}{c}{Adherence to Both Inputs}
    & \multicolumn{1}{c}{Quality} \\
        
    \cmidrule(lr){2-5}
    \cmidrule(lr){6-6}
    \cmidrule(lr){7-10}
    \cmidrule(lr){11-11}

    & Loud. $\uparrow$ & Rhyth. $\uparrow$ & Timb. $\downarrow$ & Harm. $\downarrow$ & FAD $\downarrow$ & Loud. $\uparrow$ & Rhyth. $\uparrow$ & Timb. $\downarrow$ & Harm. $\downarrow$ & FAD $\downarrow$\\
    
    \midrule
    
    Masked Tok. & - & - & - & - & 4.317 & - & - & - & - & 6.033 \\
    Guidance & 0.529 & 0.813 & 1.430 & 0.099 & 1.061 & 0.557 & 0.832 & 1.607 & 0.114 & 1.466 \\
    
    ILVR & 0.575 & 0.839 & 0.781 & 0.100  & 1.537 & 0.624 & 0.858 & \textbf{0.825} & 0.112 & 2.696 \\
    DAC 
    & 0.661 & 0.838 & 4.064 & 0.209 & 7.016 & 0.550 & 0.792 & 3.980 & 0.236 & 6.257 \\
    \Rebuttaleight{RAVE} & \Rebuttaleight{-0.016} & \Rebuttaleight{0.718} & \Rebuttaleight{3.836} & \Rebuttaleight{0.180} & \Rebuttaleight{4.695} & \Rebuttaleight{-0.006} & \Rebuttaleight{0.697} & \Rebuttaleight{4.439} & \Rebuttaleight{0.171} & \Rebuttaleight{4.478} \\
    Spectrogram & 0.366 & 0.858 & 2.104 & 0.139 & 7.608 & 0.272 & 0.824 & 2.975 & 0.128 & 7.021 \\
    Cross Synth.  & - & - & - & - & - & - & - & - & - & 2.447 \\
    \midrule
    \name-MLP & 0.815 & 0.963 & \textbf{0.376} & \textbf{0.079} & \textbf{0.337} & 0.686 & 0.873 & 1.021 & \textbf{0.108} & 1.387 \\
    \name-UNet & 0.834 & \textbf{0.966} & 0.391 & \textbf{0.079} & 0.348 & 0.686 & \textbf{0.878} & 1.118 & 0.109 & \textbf{1.357}     \\
    \Rebuttalfour{\name-DAC} & \Rebuttalfour{\textbf{0.878}} & \Rebuttalfour{0.922} & \Rebuttalfour{1.390} & \Rebuttalfour{0.107} & \Rebuttalfour{0.915} & \Rebuttalfour{\textbf{0.699}} & \Rebuttalfour{0.846} & \Rebuttalfour{1.865} & \Rebuttalfour{0.131} & \Rebuttalfour{1.364} \\


    
    \bottomrule
\end{tabular}
}%
\caption{Results on Conditional Generation and Blending on the MTG-Jamendo Test set. Mel-Cepstral Distortion (Timbre) is divided by 100.
Compared to baselines, \name variants achieve superior adherence and audio quality. The Masked Token Model and Cross Synthesis baselines do not offer frequency-based controls, so we do not compute adherence. Cross Synthesis also only applies to the blending task.}
\label{tab:condandblend}
\vspace{-12pt}
\end{table*}

\myparagraph{Results and Analysis.}
\label{sec:analysis}
We recommend listening to the qualitative results, which are available on the \href{https://masonlwang.com/latentfouriertransform/}{website}\footnote{\url{https://masonlwang.com/latentfouriertransform/}}, and show our variations are diverse and musically interesting. 
Quantitative results are in Table~\ref{tab:condandblend}. Our model outperforms all baselines in terms of adherence, indicating that the latent spectrum captures and reproduces variations in loudness, rhythm, timbre, and harmony occurring at selected timescales. 
We also surpass all baselines in terms of quality. Our metrics confirm that (1) previous audio \emph{generation} models cannot condition on features from arbitrary timescales, and (2) previous \emph{representations} of audio are not robust to post-hoc spectral modifications.


\subsection{Blending}
\label{sec:blend}
\myparagraph{Setup.}
We show that \name can blend two songs together, while preserving patterns from each at user-specified latent frequencies. This application is motivated by the traditional equalizer, whose primary use is to promote coherence between tracks by emphasizing different audible frequencies from each of them. The experimental setup is similar to the conditional generation experiment. However, instead of selecting a single latent frequency band from a single song, we select two non-overlapping bands from two songs (details in Appendix~\ref{app:condandblend}). We then measure the blended song's adherence to each song with respect to its selected subband, and average the two. To ensure that the blending is successful and musically coherent, we also report the FAD. 



\myparagraph{Analysis.} 
We provide examples of blending on the \href{https://masonlwang.com/latentfouriertransform/}{website}, and quantitative results are shown in Table~\ref{tab:condandblend}. The blending task requires an adherence--quality tradeoff, since adhering to both conditions perfectly may not result in pleasant audio. Since ILVR iteratively \emph{replaces} frequency components of the output with those of the conditions, it has a slightly better adherence score on the timbre metric, while being worse in terms of quality. ILVR also loses to \name in user studies by a substantial margin (Fig.~\ref{fig:wins}) in terms of both audio quality and ability to blend. In general, \name can better adhere to two conditions simultaneously compared to baselines, and generates higher-quality audio. The ability to adhere to disjoint latent-frequency components from two reference examples also indicates that the latent spectrum separates information by timescale to some extent.
\subsection{Listening Study}
\label{sec:listening}
\begin{figure}
    \centering
    \begin{minipage}[t]{0.3 \linewidth}
        \centering
        \includegraphics[width=\linewidth]{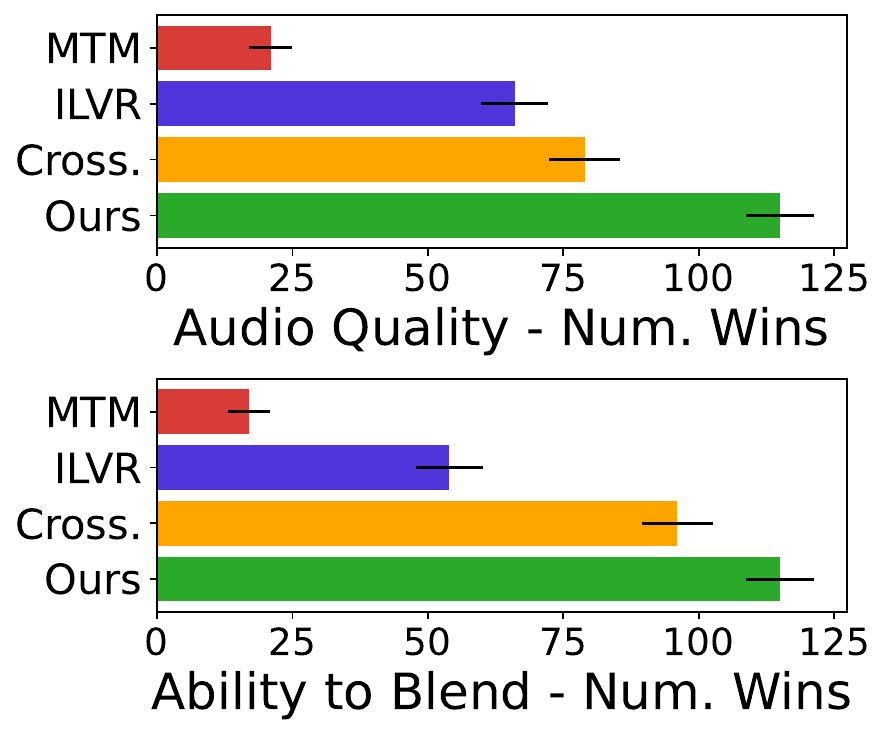}
        \caption{Listening study with pairwise comparisons. We achieve the most head-to-head wins on both criteria.}
        \label{fig:wins}
        \vspace{-7pt}
    \end{minipage}
    \hfill
    \centering
    \begin{minipage}[t]{0.69 \linewidth}
        \centering
        \begin{tikzpicture}
            \node[anchor=south west, inner sep=0] (image) at (0,0){\includegraphics[width=\linewidth]{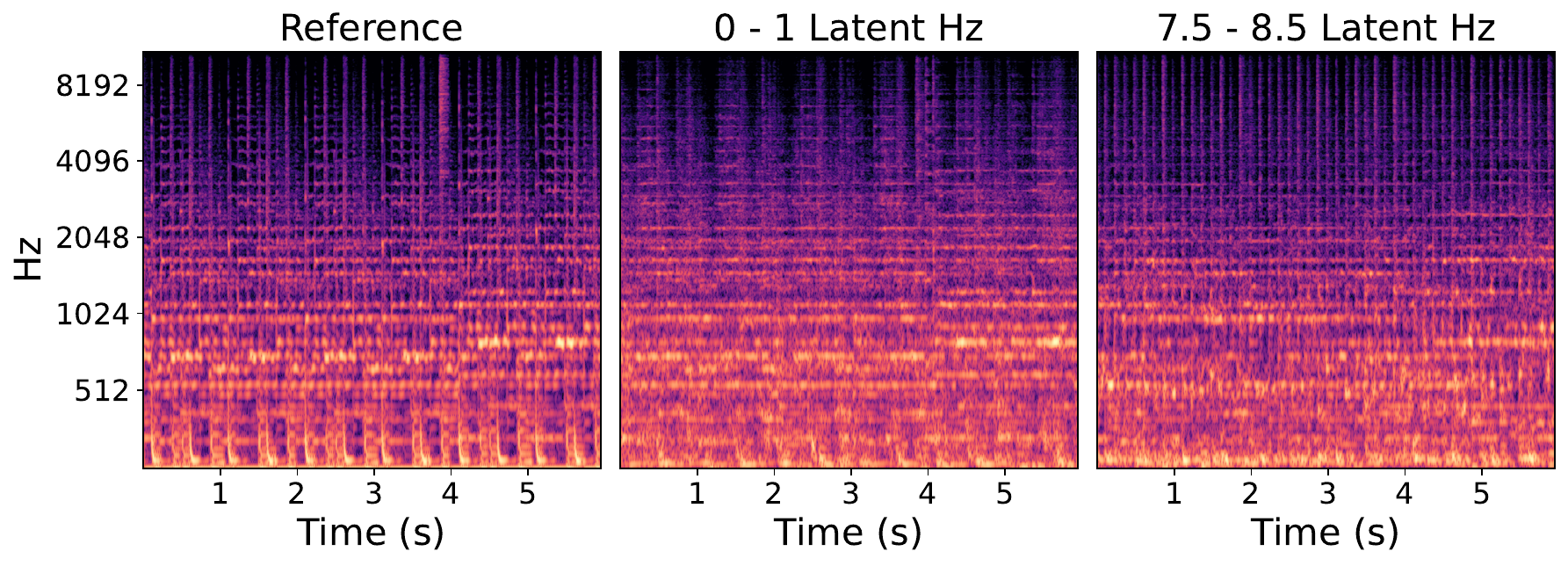}};
            \begin{scope}[x={(image.south east)}, y={(image.north west)}]
            \draw[->, black, thick] (0.345, 0.1) -- (0.345, 0.175);
            \node[below] at (0.345, 0.12) {\scriptsize Bass};
            \draw[->, black, thick] (0.693, 0.1) -- (0.653, 0.175);
            \draw[->, black, thick] (0.693, 0.1) -- (0.733, 0.175);
            \node[below] at (0.693, 0.12) {\scriptsize Bass Reduced};
            \draw[->, white] (0.8515, 0.68) -- (0.84, 0.75);
            \draw[->, white] (0.8515, 0.68) -- (0.863, 0.75);
            \node[below, white] at (0.85, 0.71) {\scriptsize 8 Hz Accentuated};
            \end{scope}
        \end{tikzpicture}
        \caption{Isolating frequencies from an electronic music clip. \Rebuttaltwo{We show three audio spectrograms}. The second spectrogram smooths the reference spectrogram, and the third accentuates patterns occurring at 8 Hz while removing lower-frequency patterns, like the bass.}
        \label{fig:iso}
        \vspace{-7pt}
    \end{minipage}
\end{figure}

To validate our method against human preferences, we conduct a listening study comparing \name and three other systems on the blending task. We choose a discrete method (the Masked Token Model baseline), a diffusion-based method (ILVR), and a traditional method (Cross Synthesis) to compare with \name. We recruited 29 musicians to complete a 12-question survey comparing every ordered pair of systems. For each question, participants first heard two randomly-selected music clips from the MTG-Jamendo test set. They then heard two blendings of the \Rebuttalfour{music} clips, each produced by a different system. Participants rated which blending they preferred in terms of (i) audio quality and (ii) how well the clips were merged, using two separate 5-point Likert scales. Fig.~\ref{fig:wins} shows that our model outperforms the baselines on both metrics. \Rebuttalfour{Additional details about the listening study and statistical analyses of the results can be found in Appendix~\ref{app:listening}.}


\subsection{Hearing in Latent Frequencies in Isolation}
\label{sec:iso}
\name can `zoom in' or `boost' patterns at specific \Rebuttaltwo{latent frequencies}, analogous to how audio engineers 
boost various audible frequencies to identify interesting or problematic regions~\citep[p.~265]{izhaki2017mixing}. 
We show this in Fig.~\ref{fig:iso}. The first spectrogram shows an electronic music clip,
containing patterns at various timescales. The second spectrogram boosts latent frequencies between 0 and 1 Hz, which removes rapid drum patterns (vertical lines near the top of the spectrograms) and bass patterns (near the bottom), 
and makes the spectrogram notably smoother along the horizontal (time) axis. The third spectrogram boosts latent frequencies between 7.5 and 8.5 Hz. This accentuates a pattern in the original song occurring at 8 Hz, seen by comparing the vertical lines in the third spectrogram with those in the first. Also, the third spectrogram \emph{does not} retain the rhythmic patterns of the bass, which occur \emph{below} 7.5 Hz. This can be seen by comparing the lower regions of spectrograms one and three. 
\name allows for performing low-pass and high-pass operations on music representations \emph{while retaining musical coherence}. Low-passing or high-passing spectrograms directly along their time axes cannot do this (Table~\ref{tab:condandblend}). We achieve isolation using a self-blending procedure described in Appendix~\ref{app:iso}.


\subsection{Interpreting the Latent Spectrum}
\label{sec:interpretability}
\begin{wrapfigure}{r}{0.35\linewidth}
    \vspace{-9pt}
    \centering
    \begin{overpic}[width=\linewidth]{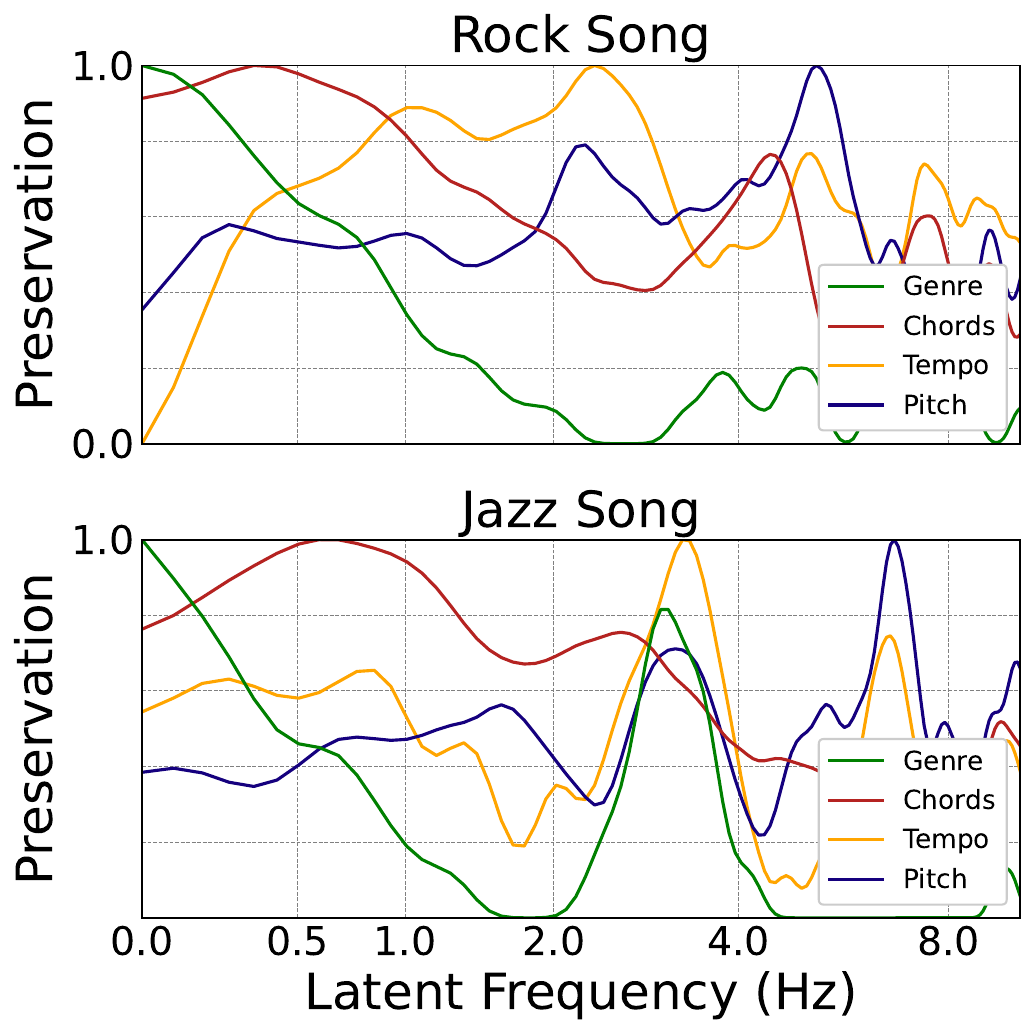}
    \end{overpic}
    \caption{Preservation curves indicating where tempo, pitch, genre reside in in the latent spectra of two reference songs.}
    \label{fig:sweep}
\end{wrapfigure}
Musical concepts like genre, tempo, pitch, and chord changes are distributed across different regions of a song's latent spectrum, analogous to how different sonic characteristics 
occupy distinct ranges of the audible spectrum. Given a song, we generate many variants while performing a sweep through the frequencies we condition on. 
For each variant, we measure preservation of genre (using a classifier), chord progression, predominant pitch, and tempo, with respect to the original song. 
We plot how well the variation preserves these traits against the frequency we condition on, applying smoothing.
Fig.~\ref{fig:sweep} shows these traits are distributed across the latent spectrum differently. Genre is a more global feature; chords change at \Rebuttaltwo{latent frequencies} below 1 Hz; and predominant pitch and tempo reside at higher frequencies, tending to be multiples of the song's BPM. For this experiment, we use the GTZAN \citep{tzanetakis2002musical} dataset, since it contains ground-truth genre labels. More details about how these preservation curves computed are in Appendix~\ref{app:sweep}. \Rebuttaltwo{Also, we interpret the latent spectra of more songs of various styles in Appendix~\ref{app:more-interpretability}.}



\section{Conclusion}
\label{sec:conclusion}
In this work, we introduced the Latent Fourier Transform, which provides novel frequency-based controls for generative models. We showed applications in conditional generation and blending in the domain of music. Future work should include enabling real-time interactivity, or disentangling the latent spectrum along semantic axes, combining both timescale-based and semantic controls.

\clearpage
\section*{Reproducibility Statement}
To promote reproducibility, we include \href{https://github.com/maswang32/latentfouriertransform/}{code}\footnote{\url{https://github.com/maswang32/latentfouriertransform/}} for training, generating, and blending examples from \name. We also include all our baseline implementations and our experiment pipeline for the conditional generation task, the blending task, code for generating sweeps for the interpretability experiment (Sec.~\ref{sec:interpretability}), and code for isolating frequency components (Sec.~\ref{sec:iso}). We also include all our model architectures, training configurations and hyperparameters (Appendix~\ref{app:details}), and code for replicating the model architecture. Finally, we include code for preprocessing the datasets we used.

\bibliography{iclr2026/iclr2026_conference}

@inproceedings{wu2018group,
  title={Group normalization},
  author={Wu, Yuxin and He, Kaiming},
  booktitle={Proceedings of the European conference on computer vision (ECCV)},
  pages={3--19},
  year={2018}
}

@inproceedings{he2016deep,
  title={Deep residual learning for image recognition},
  author={He, Kaiming and Zhang, Xiangyu and Ren, Shaoqing and Sun, Jian},
  booktitle={Proceedings of the IEEE conference on computer vision and pattern recognition},
  pages={770--778},
  year={2016}
}

@article{kong2020diffwave,
  title={Diffwave: A versatile diffusion model for audio synthesis},
  author={Kong, Zhifeng and Ping, Wei and Huang, Jiaji and Zhao, Kexin and Catanzaro, Bryan},
  journal={arXiv preprint arXiv:2009.09761},
  year={2020}
}

@article{liu2023audioldm,
  title={Audioldm: Text-to-audio generation with latent diffusion models},
  author={Liu, Haohe and Chen, Zehua and Yuan, Yi and Mei, Xinhao and Liu, Xubo and Mandic, Danilo and Wang, Wenwu and Plumbley, Mark D},
  journal={arXiv preprint arXiv:2301.12503},
  year={2023}
}

@article{huang2023noise2music,
  title={Noise2music: Text-conditioned music generation with diffusion models},
  author={Huang, Qingqing and Park, Daniel S and Wang, Tao and Denk, Timo I and Ly, Andy and Chen, Nanxin and Zhang, Zhengdong and Zhang, Zhishuai and Yu, Jiahui and Frank, Christian and others},
  journal={arXiv preprint arXiv:2302.03917},
  year={2023}
}

@article{borsos2023audiolm,
  title={Audiolm: a language modeling approach to audio generation},
  author={Borsos, Zal{\'a}n and Marinier, Rapha{\"e}l and Vincent, Damien and Kharitonov, Eugene and Pietquin, Olivier and Sharifi, Matt and Roblek, Dominik and Teboul, Olivier and Grangier, David and Tagliasacchi, Marco and others},
  journal={IEEE/ACM transactions on audio, speech, and language processing},
  volume={31},
  pages={2523--2533},
  year={2023},
  publisher={IEEE}
}

@article{agostinelli2023musiclm,
  title={Musiclm: Generating music from text},
  author={Agostinelli, Andrea and Denk, Timo I and Borsos, Zal{\'a}n and Engel, Jesse and Verzetti, Mauro and Caillon, Antoine and Huang, Qingqing and Jansen, Aren and Roberts, Adam and Tagliasacchi, Marco and others},
  journal={arXiv preprint arXiv:2301.11325},
  year={2023}
}

@article{garcia2023vampnet,
  title={Vampnet: Music generation via masked acoustic token modeling},
  author={Garcia, Hugo Flores and Seetharaman, Prem and Kumar, Rithesh and Pardo, Bryan},
  journal={arXiv preprint arXiv:2307.04686},
  year={2023}
}

@article{yang2023diffsound,
  title={Diffsound: Discrete diffusion model for text-to-sound generation},
  author={Yang, Dongchao and Yu, Jianwei and Wang, Helin and Wang, Wen and Weng, Chao and Zou, Yuexian and Yu, Dong},
  journal={IEEE/ACM Transactions on Audio, Speech, and Language Processing},
  volume={31},
  pages={1720--1733},
  year={2023},
  publisher={IEEE}
}

@article{qiu2024efficient,
  title={Efficient autoregressive audio modeling via next-scale prediction},
  author={Qiu, Kai and Li, Xiang and Chen, Hao and Sun, Jie and Wang, Jinglu and Lin, Zhe and Savvides, Marios and Raj, Bhiksha},
  journal={arXiv preprint arXiv:2408.09027},
  year={2024}
}

@article{meng2021sdedit,
  title={Sdedit: Guided image synthesis and editing with stochastic differential equations},
  author={Meng, Chenlin and He, Yutong and Song, Yang and Song, Jiaming and Wu, Jiajun and Zhu, Jun-Yan and Ermon, Stefano},
  journal={arXiv preprint arXiv:2108.01073},
  year={2021}
}

@article{choi2021ilvr,
  title={Ilvr: Conditioning method for denoising diffusion probabilistic models},
  author={Choi, Jooyoung and Kim, Sungwon and Jeong, Yonghyun and Gwon, Youngjune and Yoon, Sungroh},
  journal={arXiv preprint arXiv:2108.02938},
  year={2021}
}

@article{zeghidour2021soundstream,
  title={Soundstream: An end-to-end neural audio codec},
  author={Zeghidour, Neil and Luebs, Alejandro and Omran, Ahmed and Skoglund, Jan and Tagliasacchi, Marco},
  journal={IEEE/ACM Transactions on Audio, Speech, and Language Processing},
  volume={30},
  pages={495--507},
  year={2021},
  publisher={IEEE}
}

@article{kumar2023high,
  title={High-fidelity audio compression with improved rvqgan},
  author={Kumar, Rithesh and Seetharaman, Prem and Luebs, Alejandro and Kumar, Ishaan and Kumar, Kundan},
  journal={Advances in Neural Information Processing Systems},
  volume={36},
  pages={27980--27993},
  year={2023}
}

@inproceedings{villa2025subtractive,
  title={Subtractive training for music stem insertion using latent diffusion models},
  author={Villa-Renteria, Ivan and Wang, Mason Long and Shah, Zachary and Li, Zhe and Kim, Soohyun and Ramachandran, Neelesh and Pilanci, Mert},
  booktitle={ICASSP 2025-2025 IEEE International Conference on Acoustics, Speech and Signal Processing (ICASSP)},
  pages={1--5},
  year={2025},
  organization={IEEE}
}

@inproceedings{garcia2025sketch2sound,
  title={Sketch2sound: Controllable audio generation via time-varying signals and sonic imitations},
  author={Garc{\'\i}a, Hugo Flores and Nieto, Oriol and Salamon, Justin and Pardo, Bryan and Seetharaman, Prem},
  booktitle={ICASSP 2025-2025 IEEE International Conference on Acoustics, Speech and Signal Processing (ICASSP)},
  pages={1--5},
  year={2025},
  organization={IEEE}
}

@inproceedings{preechakul2022diffusion,
  title={Diffusion autoencoders: Toward a meaningful and decodable representation},
  author={Preechakul, Konpat and Chatthee, Nattanat and Wizadwongsa, Suttisak and Suwajanakorn, Supasorn},
  booktitle={Proceedings of the IEEE/CVF conference on computer vision and pattern recognition},
  pages={10619--10629},
  year={2022}
}

@book{izhaki2017mixing,
  title={Mixing audio: concepts, practices, and tools},
  author={Izhaki, Roey},
  year={2017},
  publisher={Routledge},
  pages={223--235},
}

@book{owsinski2017mixing,
  title={The mixing engineer's handbook},
  author={Owsinski, Bobby},
  year={2017},
  publisher={BOMG Publishing Burbank, CA},

}

@inproceedings{sohl2015deep,
  title={Deep unsupervised learning using nonequilibrium thermodynamics},
  author={Sohl-Dickstein, Jascha and Weiss, Eric and Maheswaranathan, Niru and Ganguli, Surya},
  booktitle={International conference on machine learning},
  pages={2256--2265},
  year={2015},
  organization={pmlr}
}

@article{song2020score,
  title={Score-based generative modeling through stochastic differential equations},
  author={Song, Yang and Sohl-Dickstein, Jascha and Kingma, Diederik P and Kumar, Abhishek and Ermon, Stefano and Poole, Ben},
  journal={arXiv preprint arXiv:2011.13456},
  year={2020}
}

@article{ho2020denoising,
  title={Denoising diffusion probabilistic models},
  author={Ho, Jonathan and Jain, Ajay and Abbeel, Pieter},
  journal={Advances in neural information processing systems},
  volume={33},
  pages={6840--6851},
  year={2020}
}

@misc{Forsgren_Martiros_2022,
  author = {Forsgren, Seth and Martiros, Hayk},
  title = {{Riffusion - Stable diffusion for real-time music generation}},
  url = {https://riffusion.com/about},
  year = {2022}
}

@inproceedings{chen2024musicldm,
  title={Musicldm: Enhancing novelty in text-to-music generation using beat-synchronous mixup strategies},
  author={Chen, Ke and Wu, Yusong and Liu, Haohe and Nezhurina, Marianna and Berg-Kirkpatrick, Taylor and Dubnov, Shlomo},
  booktitle={ICASSP 2024-2024 IEEE International Conference on Acoustics, Speech and Signal Processing (ICASSP)},
  pages={1206--1210},
  year={2024},
  organization={IEEE}
}

@inproceedings{schneider2024mousai,
  title={Mo{\^u}sai: Efficient text-to-music diffusion models},
  author={Schneider, Flavio and Kamal, Ojasv and Jin, Zhijing and Sch{\"o}lkopf, Bernhard},
  booktitle={Proceedings of the 62nd Annual Meeting of the Association for Computational Linguistics (Volume 1: Long Papers)},
  pages={8050--8068},
  year={2024}
}

@inproceedings{evans2025stable,
  title={Stable audio open},
  author={Evans, Zach and Parker, Julian D and Carr, CJ and Zukowski, Zack and Taylor, Josiah and Pons, Jordi},
  booktitle={ICASSP 2025-2025 IEEE International Conference on Acoustics, Speech and Signal Processing (ICASSP)},
  pages={1--5},
  year={2025},
  organization={IEEE}
}

@article{copet2023simple,
  title={Simple and controllable music generation},
  author={Copet, Jade and Kreuk, Felix and Gat, Itai and Remez, Tal and Kant, David and Synnaeve, Gabriel and Adi, Yossi and D{\'e}fossez, Alexandre},
  journal={Advances in Neural Information Processing Systems},
  volume={36},
  pages={47704--47720},
  year={2023}
}

@article{borsos2023soundstorm,
  title={Soundstorm: Efficient parallel audio generation},
  author={Borsos, Zal{\'a}n and Sharifi, Matt and Vincent, Damien and Kharitonov, Eugene and Zeghidour, Neil and Tagliasacchi, Marco},
  journal={arXiv preprint arXiv:2305.09636},
  year={2023}
}

@article{wu2024music,
  title={Music controlnet: Multiple time-varying controls for music generation},
  author={Wu, Shih-Lun and Donahue, Chris and Watanabe, Shinji and Bryan, Nicholas J},
  journal={IEEE/ACM Transactions on Audio, Speech, and Language Processing},
  volume={32},
  pages={2692--2703},
  year={2024},
  publisher={IEEE}
}

@inproceedings{parker2024stemgen,
  title={Stemgen: A music generation model that listens},
  author={Parker, Julian D and Spijkervet, Janne and Kosta, Katerina and Yesiler, Furkan and Kuznetsov, Boris and Wang, Ju-Chiang and Avent, Matt and Chen, Jitong and Le, Duc},
  booktitle={ICASSP 2024-2024 IEEE International Conference on Acoustics, Speech and Signal Processing (ICASSP)},
  pages={1116--1120},
  year={2024},
  organization={IEEE}
}

@article{novack2024ditto,
  title={Ditto: Diffusion inference-time t-optimization for music generation},
  author={Novack, Zachary and McAuley, Julian and Berg-Kirkpatrick, Taylor and Bryan, Nicholas J},
  journal={arXiv preprint arXiv:2401.12179},
  year={2024}
}

@article{levy2023controllable ,
  title={Controllable music production with diffusion models and guidance gradients},
  author={Levy, Mark and Di Giorgi, Bruno and Weers, Floris and Katharopoulos, Angelos and Nickson, Tom},
  journal={arXiv preprint arXiv:2311.00613},
  year={2023}
}

@article{lee2021fnet,
  title={Fnet: Mixing tokens with fourier transforms},
  author={Lee-Thorp, James and Ainslie, Joshua and Eckstein, Ilya and Ontanon, Santiago},
  journal={arXiv preprint arXiv:2105.03824},
  year={2021}
}

@article{he2023fourier,
  title={Fourier transformer: Fast long range modeling by removing sequence redundancy with fft operator},
  author={He, Ziwei and Yang, Meng and Feng, Minwei and Yin, Jingcheng and Wang, Xinbing and Leng, Jingwen and Lin, Zhouhan},
  journal={arXiv preprint arXiv:2305.15099},
  year={2023}
}

@article{rao2021global,
  title={Global filter networks for image classification},
  author={Rao, Yongming and Zhao, Wenliang and Zhu, Zheng and Lu, Jiwen and Zhou, Jie},
  journal={Advances in neural information processing systems},
  volume={34},
  pages={980--993},
  year={2021}
}

@article{chi2020fast,
  title={Fast fourier convolution},
  author={Chi, Lu and Jiang, Borui and Mu, Yadong},
  journal={Advances in Neural Information Processing Systems},
  volume={33},
  pages={4479--4488},
  year={2020}
}

@article{guibas2021adaptive,
  title={Adaptive fourier neural operators: Efficient token mixers for transformers},
  author={Guibas, John and Mardani, Morteza and Li, Zongyi and Tao, Andrew and Anandkumar, Anima and Catanzaro, Bryan},
  journal={arXiv preprint arXiv:2111.13587},
  year={2021}
}

@inproceedings{lin2023deep,
  title={Deep frequency filtering for domain generalization},
  author={Lin, Shiqi and Zhang, Zhizheng and Huang, Zhipeng and Lu, Yan and Lan, Cuiling and Chu, Peng and You, Quanzeng and Wang, Jiang and Liu, Zicheng and Parulkar, Amey and others},
  booktitle={Proceedings of the IEEE/CVF conference on computer vision and pattern recognition},
  pages={11797--11807},
  year={2023}
}

@article{mathieu2013fast,
  title={Fast training of convolutional networks through ffts},
  author={Mathieu, Michael and Henaff, Mikael and LeCun, Yann},
  journal={arXiv preprint arXiv:1312.5851},
  year={2013}
}

@inproceedings{ding2017circnn,
  title={Circnn: accelerating and compressing deep neural networks using block-circulant weight matrices},
  author={Ding, Caiwen and Liao, Siyu and Wang, Yanzhi and Li, Zhe and Liu, Ning and Zhuo, Youwei and Wang, Chao and Qian, Xuehai and Bai, Yu and Yuan, Geng and others},
  booktitle={Proceedings of the 50th Annual IEEE/ACM International Symposium on Microarchitecture},
  pages={395--408},
  year={2017}
}

@article{tamkin2020language,
  title={Language through a prism: A spectral approach for multiscale language representations},
  author={Tamkin, Alex and Jurafsky, Dan and Goodman, Noah},
  journal={Advances in Neural Information Processing Systems},
  volume={33},
  pages={5492--5504},
  year={2020}
}

@inproceedings{khan2017scene,
  title={Scene categorization with spectral features},
  author={Khan, Salman H and Hayat, Munawar and Porikli, Fatih},
  booktitle={Proceedings of the IEEE international conference on computer vision},
  pages={5638--5648},
  year={2017}
}

@inproceedings{yang2020fda,
  title={Fda: Fourier domain adaptation for semantic segmentation},
  author={Yang, Yanchao and Soatto, Stefano},
  booktitle={Proceedings of the IEEE/CVF conference on computer vision and pattern recognition},
  pages={4085--4095},
  year={2020}
}

@inproceedings{lee2018single,
  title={Single-image depth estimation based on fourier domain analysis},
  author={Lee, Jae-Han and Heo, Minhyeok and Kim, Kyung-Rae and Kim, Chang-Su},
  booktitle={Proceedings of the IEEE conference on computer vision and pattern recognition},
  pages={330--339},
  year={2018}
}

@article{atzmon2024edify,
  title={Edify image: High-quality image generation with pixel space laplacian diffusion models},
  author={Atzmon, Yuval and Bala, Maciej and Balaji, Yogesh and Cai, Tiffany and Cui, Yin and Fan, Jiaojiao and Ge, Yunhao and Gururani, Siddharth and Huffman, Jacob and Isaac, Ronald and others},
  journal={arXiv preprint arXiv:2411.07126},
  year={2024}
}

@article{zheng2024mfae,
  title={Mfae: Masked frequency autoencoders for domain generalization face anti-spoofing},
  author={Zheng, Tianyi and Li, Bo and Wu, Shuang and Wan, Ben and Mu, Guodong and Liu, Shice and Ding, Shouhong and Wang, Jia},
  journal={IEEE transactions on information forensics and security},
  volume={19},
  pages={4058--4069},
  year={2024},
  publisher={IEEE}
}

@inproceedings{he2022masked,
  title={Masked autoencoders are scalable vision learners},
  author={He, Kaiming and Chen, Xinlei and Xie, Saining and Li, Yanghao and Doll{\'a}r, Piotr and Girshick, Ross},
  booktitle={Proceedings of the IEEE/CVF conference on computer vision and pattern recognition},
  pages={16000--16009},
  year={2022}
}

@inproceedings{xie2022simmim,
  title={Simmim: A simple framework for masked image modeling},
  author={Xie, Zhenda and Zhang, Zheng and Cao, Yue and Lin, Yutong and Bao, Jianmin and Yao, Zhuliang and Dai, Qi and Hu, Han},
  booktitle={Proceedings of the IEEE/CVF conference on computer vision and pattern recognition},
  pages={9653--9663},
  year={2022}
}

@article{moliner2024diffusion,
  title={A diffusion-based generative equalizer for music restoration},
  author={Moliner, Eloi and Turunen, Maija and Elvander, Filip and V{\"a}lim{\"a}ki, Vesa},
  journal={arXiv preprint arXiv:2403.18636},
  year={2024}
}

@article{san2023discrete,
  title={From discrete tokens to high-fidelity audio using multi-band diffusion},
  author={San Roman, Robin and Adi, Yossi and Deleforge, Antoine and Serizel, Romain and Synnaeve, Gabriel and D{\'e}fossez, Alexandre},
  journal={Advances in Neural Information Processing Systems},
  volume={36},
  pages={1526--1538},
  year={2023}
}

@article{ashikhmin2003fast,
  title={Fast texture transfer},
  author={Ashikhmin, N},
  journal={IEEE computer Graphics and Applications},
  volume={23},
  number={4},
  pages={38--43},
  year={2003},
  publisher={IEEE}
}

@inproceedings{gatys2016image,
  title={Image style transfer using convolutional neural networks},
  author={Gatys, Leon A and Ecker, Alexander S and Bethge, Matthias},
  booktitle={Proceedings of the IEEE conference on computer vision and pattern recognition},
  pages={2414--2423},
  year={2016}
}

@inproceedings{johnson2016perceptual,
  title={Perceptual losses for real-time style transfer and super-resolution},
  author={Johnson, Justin and Alahi, Alexandre and Fei-Fei, Li},
  booktitle={European conference on computer vision},
  pages={694--711},
  year={2016},
  organization={Springer}
}

@inproceedings{huang2017arbitrary,
  title={Arbitrary style transfer in real-time with adaptive instance normalization},
  author={Huang, Xun and Belongie, Serge},
  booktitle={Proceedings of the IEEE international conference on computer vision},
  pages={1501--1510},
  year={2017}
}

@inproceedings{deng2022stytr2,
  title={Stytr2: Image style transfer with transformers},
  author={Deng, Yingying and Tang, Fan and Dong, Weiming and Ma, Chongyang and Pan, Xingjia and Wang, Lei and Xu, Changsheng},
  booktitle={Proceedings of the IEEE/CVF conference on computer vision and pattern recognition},
  pages={11326--11336},
  year={2022}
}

@incollection{efros2023image,
  title={Image quilting for texture synthesis and transfer},
  author={Efros, Alexei A and Freeman, William T},
  booktitle={Seminal graphics papers: pushing the boundaries, volume 2},
  pages={571--576},
  year={2023}
}

@article{dai2018music,
  title={Music style transfer: A position paper},
  author={Dai, Shuqi and Zhang, Zheng and Xia, Gus G},
  journal={arXiv preprint arXiv:1803.06841},
  year={2018}
}

@article{huang2018timbretron,
  title={Timbretron: A wavenet (cyclegan (cqt (audio))) pipeline for musical timbre transfer},
  author={Huang, Sicong and Li, Qiyang and Anil, Cem and Bao, Xuchan and Oore, Sageev and Grosse, Roger B},
  journal={arXiv preprint arXiv:1811.09620},
  year={2018}
}

@article{wang2024training,
  title={A training-free approach for music style transfer with latent diffusion models},
  author={Wang, Heehwan and Kwon, Joonwoo and Kim, Sooyoung and Yoo, Shinjae and Lin, Yuewei and Cha, Jiook},
  journal={arXiv preprint arXiv:2411.15913},
  year={2024}
}

@inproceedings{li2024music,
  title={Music style transfer with time-varying inversion of diffusion models},
  author={Li, Sifei and Zhang, Yuxin and Tang, Fan and Ma, Chongyang and Dong, Weiming and Xu, Changsheng},
  booktitle={Proceedings of the AAAI Conference on Artificial Intelligence},
  volume={38},
  number={1},
  pages={547--555},
  year={2024}
}

@article{cifka2020groove2groove,
  title={Groove2groove: One-shot music style transfer with supervision from synthetic data},
  author={C{\'\i}fka, Ond{\v{r}}ej and {\c{S}}im{\c{s}}ekli, Umut and Richard, Ga{\"e}l},
  journal={IEEE/ACM Transactions on Audio, Speech, and Language Processing},
  volume={28},
  pages={2638--2650},
  year={2020},
  publisher={IEEE}
}

@inproceedings{se2016musical,
  title={Musical style modification as an optimization problem},
  author={SE, SAP},
  booktitle={Proceedings of the International Computer Music Conference},
  pages={206},
  year={2016}
}

@book{smith2011spectral,
  title={Spectral audio signal processing},
  author={Smith, Julius O},
  publisher={Julius Smith},
  year={2011}
}

@article{lee2022bigvgan,
  title={Bigvgan: A universal neural vocoder with large-scale training},
  author={Lee, Sang-gil and Ping, Wei and Ginsburg, Boris and Catanzaro, Bryan and Yoon, Sungroh},
  journal={arXiv preprint arXiv:2206.04658},
  year={2022}
}

@article{sonderby2016ladder,
  title={Ladder variational autoencoders},
  author={S{\o}nderby, Casper Kaae and Raiko, Tapani and Maal{\o}e, Lars and S{\o}nderby, S{\o}ren Kaae and Winther, Ole},
  journal={Advances in neural information processing systems},
  volume={29},
  year={2016}
}

@article{gulrajani2016pixelvae,
  title={Pixelvae: A latent variable model for natural images},
  author={Gulrajani, Ishaan and Kumar, Kundan and Ahmed, Faruk and Taiga, Adrien Ali and Visin, Francesco and Vazquez, David and Courville, Aaron},
  journal={arXiv preprint arXiv:1611.05013},
  year={2016}
}

@article{bachman2016architecture,
  title={An architecture for deep, hierarchical generative models},
  author={Bachman, Philip},
  journal={Advances in Neural Information Processing Systems},
  volume={29},
  year={2016}
}

@article{zhao2017learning,
  title={Learning hierarchical features from generative models},
  author={Zhao, Shengjia and Song, Jiaming and Ermon, Stefano},
  journal={arXiv preprint arXiv:1702.08396},
  year={2017}
}

@article{pasini2024music2latent,
  title={Music2latent: Consistency autoencoders for latent audio compression},
  author={Pasini, Marco and Lattner, Stefan and Fazekas, George},
  journal={arXiv preprint arXiv:2408.06500},
  year={2024}
}

@article{bindi2024unsupervised,
  title={Unsupervised composable representations for audio},
  author={Bindi, Giovanni and Esling, Philippe},
  journal={arXiv preprint arXiv:2408.09792},
  year={2024}
}

@article{stevens1937scale,
  title={A scale for the measurement of the psychological magnitude pitch},
  author={Stevens, Stanley Smith and Volkmann, John and Newman, Edwin Broomell},
  journal={The journal of the acoustical society of america},
  volume={8},
  number={3},
  pages={185--190},
  year={1937},
  publisher={Acoustical Society of America}
}

@article{brown1991calculation,
  title={Calculation of a constant Q spectral transform},
  author={Brown, Judith C},
  journal={The Journal of the Acoustical Society of America},
  volume={89},
  number={1},
  pages={425--434},
  year={1991},
  publisher={Acoustical Society of America}
}

@article{bak1987self,
  title={Self-organized criticality: An explanation of the 1/f noise},
  author={Bak, Per and Tang, Chao and Wiesenfeld, Kurt},
  journal={Physical review letters},
  volume={59},
  number={4},
  pages={381},
  year={1987},
  publisher={APS}
}

@book{smith2007mathematics,
  title={Mathematics of the discrete Fourier transform (DFT): with audio applications},
  author={Smith, Julius O},
  year={2007},
  publisher={Julius Smith}
}

@article{karras2022elucidating,
  title={Elucidating the design space of diffusion-based generative models},
  author={Karras, Tero and Aittala, Miika and Aila, Timo and Laine, Samuli},
  journal={Advances in neural information processing systems},
  volume={35},
  pages={26565--26577},
  year={2022}
}

@inproceedings{bogdanov2019mtg,
  title={The mtg-jamendo dataset for automatic music tagging},
  author={Bogdanov, Dmitry and Won, Minz and Tovstogan, Philip and Porter, Alastair and Serra, Xavier},
  year={2019},
  organization={ICML}
}

@inproceedings{hasumi2025music,
  title={Music tagging with classifier group chains},
  author={Hasumi, Takuya and Komatsu, Tatsuya and Fujita, Yusuke},
  booktitle={ICASSP 2025-2025 IEEE International Conference on Acoustics, Speech and Signal Processing (ICASSP)},
  pages={1--5},
  year={2025},
  organization={IEEE}
}

@inproceedings{lanzendorfer2025high,
  title={High-fidelity music vocoder using neural audio codecs},
  author={Lanzend{\"o}rfer, Luca A and Gr{\"o}tschla, Florian and Ungersb{\"o}ck, Michael and Wattenhofer, Roger},
  booktitle={ICASSP 2025-2025 IEEE International Conference on Acoustics, Speech and Signal Processing (ICASSP)},
  pages={1--5},
  year={2025},
  organization={IEEE}
}

@article{tzanetakis2002musical,
  title={Musical genre classification of audio signals},
  author={Tzanetakis, George and Cook, Perry},
  journal={IEEE Transactions on speech and audio processing},
  volume={10},
  number={5},
  pages={293--302},
  year={2002},
  publisher={IEEE}
}

@article{morrison2024fine,
  title={Fine-grained and interpretable neural speech editing},
  author={Morrison, Max and Churchwell, Cameron and Pruyne, Nathan and Pardo, Bryan},
  journal={arXiv preprint arXiv:2407.05471},
  year={2024}
}

@article{kosta2016mapping,
  title={Mapping between dynamic markings and performed loudness: a machine learning approach},
  author={Kosta, Katerina and Ram{\'\i}rez, Rafael and Bandtlow, Oscar F and Chew, Elaine},
  journal={Journal of Mathematics and Music},
  volume={10},
  number={2},
  pages={149--172},
  year={2016},
  publisher={Taylor \& Francis}
}

@inproceedings{kominek2008synthesizer,
  title={Synthesizer voice quality of new languages calibrated with mean mel cepstral distortion.},
  author={Kominek, John and Schultz, Tanja and Black, Alan W},
  booktitle={SLTU},
  pages={63--68},
  year={2008}
}

@inproceedings{bock2013maximum,
  title={Maximum filter vibrato suppression for onset detection},
  author={B{\"o}ck, Sebastian and Widmer, Gerhard},
  booktitle={Proc. of the 16th Int. Conf. on Digital Audio Effects (DAFx). Maynooth, Ireland (Sept 2013)},
  volume={7},
  pages={4},
  year={2013},
  organization={Citeseer}
}

@inproceedings{foote2002audio,
  title={Audio Retrieval by Rhythmic Similarity.},
  author={Foote, Jonathan and Cooper, Matthew and Nam, Unjung},
  booktitle={ISMIR},
  year={2002}
}

@article{milne2016empirically,
  title={Empirically testing Tonnetz, voice-leading, and spectral models of perceived triadic distance},
  author={Milne, Andrew J and Holland, Simon},
  journal={Journal of Mathematics and Music},
  volume={10},
  number={1},
  pages={59--85},
  year={2016},
  publisher={Taylor \& Francis}
}

@article{kilgour2018fr,
  title={Fr$\backslash$'echet audio distance: A metric for evaluating music enhancement algorithms},
  author={Kilgour, Kevin and Zuluaga, Mauricio and Roblek, Dominik and Sharifi, Matthew},
  journal={arXiv preprint arXiv:1812.08466},
  year={2018}
}

@inproceedings{ronneberger2015u,
  title={U-net: Convolutional networks for biomedical image segmentation},
  author={Ronneberger, Olaf and Fischer, Philipp and Brox, Thomas},
  booktitle={International Conference on Medical image computing and computer-assisted intervention},
  pages={234--241},
  year={2015},
  organization={Springer}
}

@article{stoller2018wave,
  title={Wave-u-net: A multi-scale neural network for end-to-end audio source separation},
  author={Stoller, Daniel and Ewert, Sebastian and Dixon, Simon},
  journal={arXiv preprint arXiv:1806.03185},
  year={2018}
}

@article{passigan2024analyzing,
  title={Analyzing the Effect of $ k $-Space Features in MRI Classification Models},
  author={Passigan, Pascal and Ramkumar, Vayd},
  journal={arXiv preprint arXiv:2409.13589},
  year={2024}
}

@inproceedings{hershey2017cnn,
  title={CNN architectures for large-scale audio classification},
  author={Hershey, Shawn and Chaudhuri, Sourish and Ellis, Daniel PW and Gemmeke, Jort F and Jansen, Aren and Moore, R Channing and Plakal, Manoj and Platt, Devin and Saurous, Rif A and Seybold, Bryan and others},
  booktitle={2017 ieee international conference on acoustics, speech and signal processing (icassp)},
  pages={131--135},
  year={2017},
  organization={IEEE}
}

@inproceedings{bogdanov2013essentia,
  title={Essentia: an open-source library for sound and music analysis},
  author={Bogdanov, Dmitry and Wack, Nicolas and G{\'o}mez, Emilia and Gulati, Sankalp and Herrera, Perfecto and Mayor, Oscar and Roma, Gerard and Salamon, Justin and Zapata, Jos{\'e} and Serra, Xavier},
  booktitle={Proceedings of the 21st ACM international conference on Multimedia},
  pages={855--858},
  year={2013}
}

@article{mcfee2015librosa,
  title={librosa: Audio and music signal analysis in python.},
  author={McFee, Brian and Raffel, Colin and Liang, Dawen and Ellis, Daniel PW and McVicar, Matt and Battenberg, Eric and Nieto, Oriol},
  journal={SciPy},
  volume={2015},
  pages={18--24},
  year={2015}
}

@article{hendrycks2016gaussian,
  title={Gaussian Error Linear Units (Gelus)},
  author={Hendrycks, D},
  journal={arXiv preprint arXiv:1606.08415},
  year={2016}
}

@article{hawthorne2018enabling,
  title={Enabling factorized piano music modeling and generation with the MAESTRO dataset},
  author={Hawthorne, Curtis and Stasyuk, Andriy and Roberts, Adam and Simon, Ian and Huang, Cheng-Zhi Anna and Dieleman, Sander and Elsen, Erich and Engel, Jesse and Eck, Douglas},
  journal={arXiv preprint arXiv:1810.12247},
  year={2018}
}

@misc{caillon2021ravevariationalautoencoderfast,
      title={RAVE: A variational autoencoder for fast and high-quality neural audio synthesis}, 
      author={Antoine Caillon and Philippe Esling},
      year={2021},
      eprint={2111.05011},
      archivePrefix={arXiv},
      primaryClass={cs.LG},
      url={https://arxiv.org/abs/2111.05011}, 
}

@article{huang2022masked,
  title={Masked autoencoders that listen},
  author={Huang, Po-Yao and Xu, Hu and Li, Juncheng and Baevski, Alexei and Auli, Michael and Galuba, Wojciech and Metze, Florian and Feichtenhofer, Christoph},
  journal={Advances in Neural Information Processing Systems},
  volume={35},
  pages={28708--28720},
  year={2022}
}

@article{landis1977measurement,
  title={The measurement of observer agreement for categorical data},
  author={Landis, J Richard and Koch, Gary G},
  journal={biometrics},
  pages={159--174},
  year={1977},
  publisher={JSTOR}
}

@inproceedings{nabi2024embodied,
  title={Embodied exploration of deep latent spaces in interactive dance-music performance},
  author={Nabi, Sarah and Esling, Philippe and Peeters, Geoffroy and Bevilacqua, Fr{\'e}d{\'e}ric},
  booktitle={Proceedings of the 9th International Conference on Movement and Computing},
  pages={1--9},
  year={2024}
}

@article{zheng2024mapping,
  title={A Mapping Strategy for Interacting with Latent Audio Synthesis Using Artistic Materials},
  author={Zheng, Shuoyang and Sed{\'o}, Anna Xamb{\'o} and Bryan-Kinns, Nick},
  journal={arXiv preprint arXiv:2407.04379},
  year={2024}
}

@article{salamon2014melody,
  title={Melody extraction from polyphonic music signals: Approaches, applications, and challenges},
  author={Salamon, Justin and G{\'o}mez, Emilia and Ellis, Daniel PW and Richard, Ga{\"e}l},
  journal={IEEE Signal Processing Magazine},
  volume={31},
  number={2},
  pages={118--134},
  year={2014},
  publisher={IEEE}
}
\bibliographystyle{iclr2026/iclr2026_conference}

\clearpage
\appendix

\etocdepthtag{appendix}

\etocsettagdepth{main}{none}

\etocsettagdepth{appendix}{subsubsection}

\etocsettocstyle{\section*{Appendix Table of Contents}}{}

{
  \hypersetup{hidelinks}
  \tableofcontents
}




\clearpage
\section{Experimental Details}
\label{app:details}
Below, we describe our experiments in more detail. We provide code for training and evaluating \name in our \href{https://github.com/maswang32/latentfouriertransform/}{public github repository}\footnote{\url{https://github.com/maswang32/latentfouriertransform/}}.

\subsection{Encoders}
\label{app:encoders}
We experiment with \Rebuttalfour{three} encoders:

\begin{enumerate}
    \item \textbf{MLP Encoder.} The audio is converted into an $80 \times 512$ mel-spectrogram. Each $(80 \times  1)$ timeframe is passed through an MLP to obtain an $80 \times 512$ latent sequence. Since each timeframe is processed independently, this encoder enforces input-output alignment, and results in no leakage between timeframes.
    \item \textbf{1D U-Net Encoder.} The audio is first converted into an $80 \times 512$ mel-spectrogram. This is processed by a 1D U-Net \citep{ronneberger2015u, stoller2018wave} with convolutions along the temporal axis to obtain an $80 \times 512$ latent sequence. While this encoder does not entirely prevent leakage between frames, the U-Net's skip-connections promote input-output alignment, allowing the encoding to be interpreted as a temporal sequence.
    \Rebuttalfour{\item \textbf{DAC Encoder.} We use the encoder from the pretrained Descript Audio Codec \citep{kumar2023high} model to extract $1024 \times 512$ embeddings from the raw audio waveform. We then pass these embeddings to a 1D U-Net encoder that is identical to the one described above, except for the first convolutional layer, which is expanded to have $1024$ input channels instead of $80$ to accommodate the number of latent channels in DAC.}
\end{enumerate}

We find qualitatively that the U-Net encoder produces more pleasant-sounding audio when listening to latent frequencies in isolation. We also find that the U-Net encoder is better for blending, while the MLP Encoder is better for conditional generation. \Rebuttalfour{The DAC encoder's waveform frontend requires significantly more GPU memory, which required reducing the batch size from 256 to 64 during training. We observe in Table~\ref{tab:condandblend} that it is better at preserving loudness curves.} \Rebuttaltwo{Below, we provide more hyperparameters for each of our encoders.}

\subsubsection{\Rebuttaltwo{MLP Encoder Architecture}}
\Rebuttaltwo{Our MLP encoder takes in an $80 \times 512$ mel-spectrogram, but processes each of the $512$ latent timeframes independently, operating only on the channel axis. It can also be thought of as a convolutional network, where the convolutions apply to the time axis and have a kernel size of 1. It consists of a series of linear layers with SiLU activations \citep{hendrycks2016gaussian}, group normalization layers \citep{wu2018group}, and residual connections \citep{he2016deep}. The hyperparameters for our MLP encoder are listed in Table~\ref{tab:mlp-params}.}

\begin{table}[h]
\centering
\small
\begin{tabular}{>{\color{black}}c >{\color{black}}c}
\toprule
\textbf{Attribute} & \textbf{Value} \\
\midrule
Input & $80 \times 512$ mel-spectrogram \\
Output & $80  \times 512$ latent sequence \\
Architecture & Frame-wise MLP \\
Hidden Dim. & 512 \\
Num. Hidden Layers & 16 \\
\bottomrule
\end{tabular}
\caption{\Rebuttaltwo{MLP Encoder Architecture}}
\label{tab:mlp-params}
\end{table}

\clearpage
\subsubsection{\Rebuttaltwo{1D U-Net Encoder Architecture}}
\label{app:unet-encoder}
\Rebuttaltwo{Our 1D U-Net encoder is a 1D version of the encoder used in \cite{karras2022elucidating}. The convolutions occur along the temporal axis. The U-Net consists of several blocks that process information at different resolutions, which are listed below in Table~\ref{tab:unet-params}. In addition, we add self-attention layers to blocks at particular resolutions, which are also listed below in Table~\ref{tab:unet-params}.}

\begin{table}[h!]
\centering
\small
\begin{tabular}{>{\color{black}}c >{\color{black}}c}
\toprule
\textbf{Attribute} & \textbf{Value} \\
\midrule
Input & $80 \times 512$ mel-spectrogram \\
Output & $80 \times 512$ latent sequence \\
Architecture & 1D U-Net \\
Kernel Size & 3 \\
Resolutions & [512, 256, 128, 64, 32, 16] \\
Channels Per Resolution & [512, 512, 512, 768, 768, 1024] \\
Resolutions with Attention & [64, 32, 16] \\
\bottomrule
\end{tabular}
\caption{\Rebuttaltwo{1D U-Net Encoder Hyperparameters}}
\label{tab:unet-params}
\end{table}

\subsubsection{\Rebuttalfour{DAC Encoder Architecture}}
\Rebuttalfour{The DAC encoder takes in a raw audio waveform which is resampled to 44.1 kHz. First, it creates a $1024 \times 512$ sequence of continuous embeddings using the encoder of Descript Audio Codec \citep{kumar2023high}. This sequence 
is passed to a 1D U-Net identical to the one in Table~\ref{tab:unet-params}, except for the first convolutional layer, which has $1024$ input channels instead of $80$, to match the latent dimension of DAC. Table~\ref{tab:dac-encoder-params} provides details.}
\begin{table}[h]
\centering
\small
\begin{tabular}{>{\color{black}}c >{\color{black}}c}
\toprule
\textbf{Attribute} & \textbf{Value} \\
\midrule
Architecture & DAC + 1D U-Net \\
DAC Encoder Input & $1 \times 262144$ audio waveform \\
DAC Encoder Output & $1024 \times 512$ DAC embedding \\ 
1D-UNet Input & $1024 \times 512$ DAC embedding \\ 
1D-UNet Output & $80 \times 512$ latent sequence \\
\bottomrule
\end{tabular}
\caption{\Rebuttalfour{DAC Encoder Hyperparameters}}
\label{tab:dac-encoder-params}
\end{table}

\subsection{\Rebuttaltwo{Decoders/Diffusion Model Architecture}}
\label{app:decoders}
\Rebuttaltwo{
Our decoder (diffusion models) are 1D-U-Nets that combine convolutional layers with self-attention layers. The decoder is very similar to the 1D-UNet encoder described in Appendix~\ref{app:unet-encoder}. The main difference is the decoder's input is a noisy mel-spectrogram $\vx_{\tau}$, as well as the masked latent $\vz_{\text{masked}}$. These two inputs are concatenated channel-wise before being fed to the U-Net. The U-Net predicts a linear combination of the added noise and the clean input $\vx_0$, as described in \cite{karras2022elucidating}. Again, we follow the architectures in \cite{karras2022elucidating} and provide our code in our github repository. Details are shown in Table~\ref{tab:decoder-params}.}

\begin{table}[h]
\centering
\small
\begin{tabular}{>{\color{black}}c >{\color{black}}c}
\toprule
\textbf{Attribute} & \textbf{Value} \\
\midrule
Input 1 & $80 \times 512$ noisy mel-spectrogram \\
Input 2 & $80 \times 512$ frequency-masked latent \\
Output & $80 \times 512$ clean mel-spectrogram \\
Architecture & 1D U-Net \\
Kernel Size & 3 \\
Resolutions & [512, 256, 128, 64, 32, 16] \\
Channels Per Resolution & [512, 512, 512, 768, 768, 1024] \\
Resolutions with Attention & [64, 32, 16] \\
\bottomrule
\end{tabular}
\caption{\Rebuttaltwo{Decoder Hyperparameters}}
\label{tab:decoder-params}
\end{table}

\subsection{\Rebuttaltwo{Training Details}}
\label{app:training}
For the experiments in the main paper, we train for 700k iterations on 4 L40S GPUs with a logical batch size of 1024. \Rebuttaltwo{We use a linear warmup for the first 4,000 training steps, and we apply cosine annealing to the learning rate after 350k iterations. In addition, following \cite{karras2022elucidating}, we store an exponential moving average of the model weights, which we use for inference. Hyperparameters for this are shown in Table~\ref{tab:training-params}. For the ablation experiments shown in Appendix~\ref{app:ablations}, we train for 350k iterations, and do not perform annealing.}

\begin{table}[h]
\centering
\small
\begin{tabular}{
    >{\color{black}}c
    >{\color{black}}c
    >{\color{black}}c
}
\toprule
 & \textbf{Attribute} & \textbf{Value} \\
\midrule

\multirow{4}{*}{\textbf{Training Schedule}}
    & Num. Total Iters   & 700k \\
    & Num. Warmup Iters  & 4k \\
    & Num. Decay Iters   & 350k \\
    & Decay Schedule     & Cosine \\

\midrule

\multirow{4}{*}{\textbf{Optimizer}}
    & Optimizer          & Adam \\
    & Learning Rate      & 1e\,-4 \\
    & $\beta_1$          & 0.9 \\
    & $\beta_2$          & 0.999 \\

\midrule

\multirow{3}{*}{\textbf{Batching}}
    & Batch Size (Logical)   & 1024 \\
    & Batch Size (Per-GPU)   & 256 \\
    & Distribution Strategy               & DDP \\

\midrule

\multirow{3}{*}{\textbf{Other}}
    & Precision          & Mixed FP32 + BF16 \\
    & Grad Clip Value   & 1.0 \\
    & EMA Decay         & 0.999 \\

\bottomrule
\end{tabular}
\caption{\Rebuttaltwo{Training Hyperparameters}}
\label{tab:training-params}
\end{table}

\subsection{\Rebuttaltwo{Other Hyperparameters}}
\Rebuttaltwo{Here, we list the values of other hyperparameters mentioned in the Methods section (Sec.~\ref{sec:method}).}
\label{app:other-hyperparameters}

\begin{table}[h]
\centering
\small
\begin{tabular}{
    >{\color{black}}c
    >{\color{black}}c
    >{\color{black}}c
}
\toprule
 & \textbf{Attribute} & \textbf{Value} \\
\midrule

\multirow{4}{*}{\textbf{DFT / Frequency Mask }}
    &  $L$ & 2\\
    &  $\sigma$ & 0.5 \\
    &  $p$& 2 \\
    &  $\epsilon $ & 1e-6 \\

\midrule
\multirow{3}{*}{\textbf{Diffusion}}
    & $\sigma_{\text{max}}$  & 80 \\
    & $\alpha$   & 0.5 \\
    & $\beta$         & 0.5 \\
\bottomrule
\end{tabular}
\caption{\Rebuttaltwo{Other Hyperparameters. Full descriptions can be found in the Methods section (Sec.~\ref{sec:method})}}
\label{tab:other-params}
\end{table}

\subsection{Datasets}
\label{app:datasets}
Our experiments in the main paper use two datasets. All clips are resampled to a sampling rate of 22050 Hz.
\begin{enumerate}
    \item \textbf{MTG-Jamendo.} MTG-Jamendo \citep{bogdanov2019mtg} is a large-scale collection of over 55,000 spanning diverse genres, like classical, electronic, pop, and rock music. The dataset is publicly available, and is popular in tasks like neural audio compression, vocoding \citep{lanzendorfer2025high}, and music-tagging \citep{hasumi2025music}. We train our models on a dataset of 2.5 million 5.9-second clips from the MTG-Jamendo training split. The MTG-Jamendo dataset is used in the conditional generation (Sec.~\ref{sec:cond}), blending (Sec.~\ref{sec:blend}), listening study (Sec.~\ref{sec:listening}), and isolation experiments (Sec.~\ref{sec:iso}).
    \item \textbf{GTZAN.} GTZAN \citep{tzanetakis2002musical} is a standard benchmark for genre classification, containing 1,000 30-second audio clips evenly distributed across 10 genres (blues, classical, country, disco, hip-hop, jazz, metal, pop, reggae, and rock). We use GTZAN for the interpretability experiment (Sec.~\ref{sec:interpretability}), since we require high-quality genre labels.
\end{enumerate}
We show results on more datasets in Appendix~\ref{app:more_datasets}, where we perform the conditional generation and blending experiments on GTZAN and the Maestro dataset \citep{hawthorne2018enabling}.


\subsection{Conditional Generation and Blending Experiments}
\label{app:condandblend}
For these experiments, we partition the latent spectrum into 2 bands, 4 bands, and 8 bands. In each of the three partitionings, bands are equal-width on a logarithmic axis. For the conditional generation task, we condition each song on all 14 bands one-at-a-time, averaging results. For the blending task, we take two examples and condition on every possible unordered pair of bands inside the 4-band partition, for a total of six possible conditions. 

\subsection{\Rebuttalfour{Listening Study Details and Analysis}}
\label{app:listening}

\Rebuttalfour{
We used \href{https://www.prolific.com/}{Prolific} to recruit high-quality participants for our survey. All respondents self-identified as musicians, and all respondents reside in the United States. The respondents ranged in age from 20--73 years old, with an average age of 43.4 years old, and a median age of 41.5 years old.
}

\Rebuttalfour{The survey consists of 12 questions, each comparing two ordered pairs of systems on the blending task. Each question presents two reference recordings, and then presents two blendings of the reference clips from two different systems, for a total of four clips. The users are asked which recording they prefer both in terms of audio quality, and how well the clips were ``blended" together.  A screenshot of a question from our survey is shown in Fig.~\ref{fig:surveyscreenshot}. 

\begin{figure}[h]
    \centering
    \includegraphics[width=\linewidth]{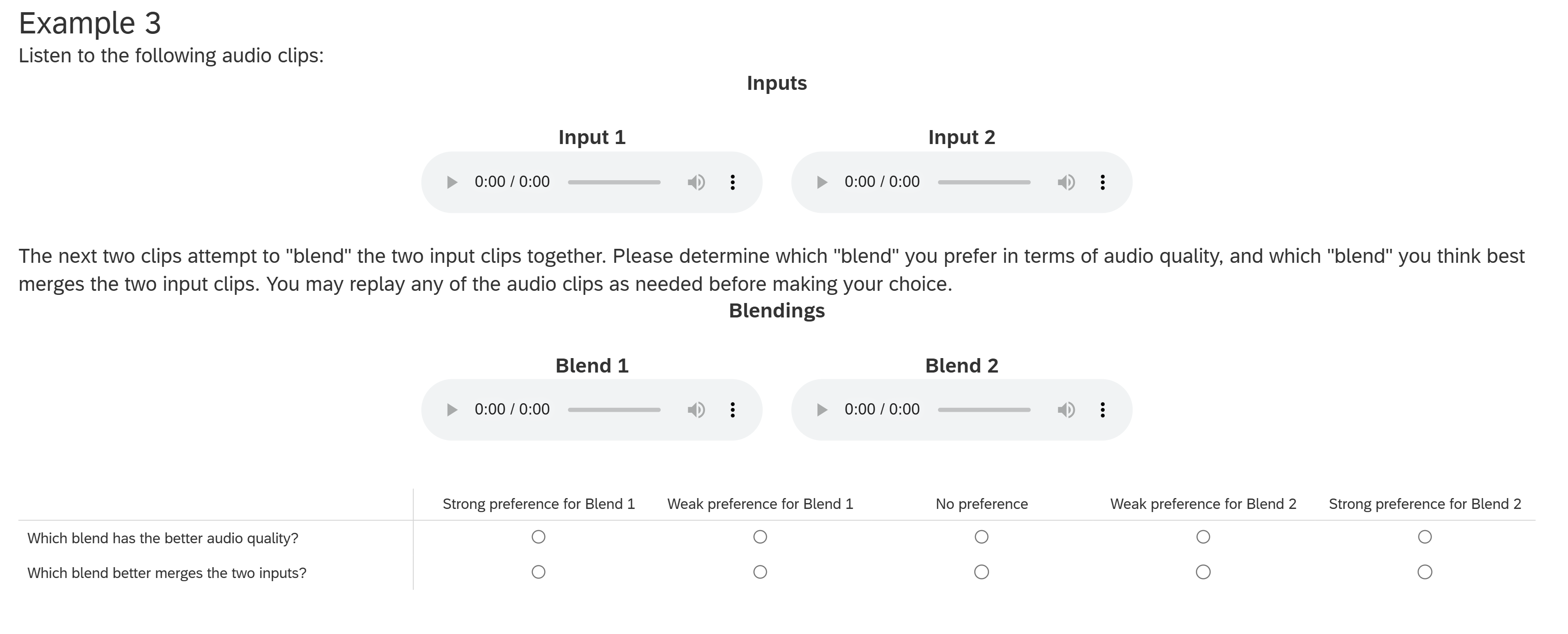}
    \caption{\Rebuttalfour{A question from our listening study survey. A participant will compare each ordered pair of systems in the study once.}}
    \label{fig:surveyscreenshot}
\end{figure}

\Rebuttalfour{The order of all questions is randomized. In addition, we include one attention check question for each survey participant. In the attention check, all recordings are silent, and the participant is instructed to select `2' and `4' for their two Likert scale ratings. The total duration of all the audio recordings in our survey was 5 minutes and 9 seconds.
However, the median survey response time was 10 minutes and 25 seconds. 
}

\myparagraph{\Rebuttalfour{Statistical Significance.}} We performed a Kruskal-Wallis H test, which confirmed that
there are statistically significant pairs among the permutations ($p=6.4 \times 10^{-83}$). We also perform a post-hoc analysis using the Wilcoxon signed rank test. We apply a Bonferroni correction, which corrects the significant threshold to $p<0.05/6$. According to this test, all pairs of systems have statistically significant differences in audio quality, except for the cross-synthesis and ILVR baselines. This means that \name outperforms all baselines in terms of audio quality according to our user study, to a statistically significant extent. Another Wilcoxon signed rank test indicates that all pairs of systems have statistically significant differences in ``ability to blend", except for \name and the cross synthesis baseline. Pairwise significance test results are shown in Table~\ref{tab:listening-statistics}.
}

\myparagraph{\Rebuttalfour{Inter-rater Agreement.}} 
\begin{table}[h]
    \centering
    \begin{tabular}{>{\color{black}}c>{\color{black}}c>{\color{black}}c>{\color{black}}c}
    \toprule
    \textbf{System 1} & \textbf{System 2} & \textbf{$p$-value, Audio Quality} & \textbf{$p$-Value, Ability to Blend} \\
    \cmidrule(lr){1-4}
    \name & Cross Synthesis & $1.59 \times 10^{-3}$ & $9.54 \times 10^{-2*}$ \\
    \name & ILVR            & $ 3.83 \times 10^{-4} $ & $8.84 \times 10^{-7}$ \\
    \name & VampNet         & $7.02 \times 10^{-10}$ & $1.64 \times 10^{-10}$ \\
    Cross Synthesis & ILVR         & $9.51 \times 10^{-2*}$ &  $6.62 \times 10^{-4}$ \\
    Cross Synthesis & VampNet         &  $1.91 \times 10^{-6}$ &  $8.09 \times 10^{-10}$\\
    ILVR & VampNet         & $1.55 \times 10^{-6}$ &  $1.69 \times 10^{-5}$ \\
    \bottomrule
    \end{tabular}
    \caption{\Rebuttalfour{Results from a Kruskal-Wallis H test performed on listening study results. All pairs of systems have statistically significant differences in audio quality, except for ILVR and Cross Synthesis. All pairs of systems have statistically significant differences in ``Ability to Blend" besides \name and Cross Synthesis. These pairs are indicated with an asterisk ($^*$).}}
    \label{tab:listening-statistics}
\end{table}
\Rebuttalfour{To compute inter-rater agreement between our 29 participants, we calculate Fleiss's Kappa, which measures the degree of agreement beyond chance for multiple raters. We report $\kappa = 0.0654$ for our question about audio quality, and $\kappa = 0.0914$ for our question about ``ability to blend". Both values fall in the ``slight agreement" range \citep{landis1977measurement}, indicating substantial subjective variation in perceptual judgments. This level of agreement is possible due to individual preferences and perceptual differences, which naturally lead to varied responses.}

\subsection{Isolation Experiments}
\label{app:iso}
We accomplish isolation by taking a music clip $\vx$ and obtaining $\vz$, the full-spectrum latent sequence, and $\vz^{\text{bp}}$, a version of the latent sequence $\vz$ bandpassed to the selected frequency range. We then guide the diffusion process with both $\vz$ and $\vz^{\text{bp}}$ (see Alg.~\ref{alg:blending}), with blend weights $\alpha, \beta$, resulting in an output that emphasizes the selected band while suppressing content outside of it. The ratio of $\beta$ and $\alpha$ determines the amount of boosting that occurs, with $\beta \gg \alpha$ resulting in isolating the selected band almost completely. 

\subsection{Interpreting the Latent Spectrum.}
\label{app:sweep}
\Rebuttaltwo{
In the interpretability experiment (Sec.~\ref{sec:interpretability}), we analyze the latent spectrum of individual songs, and associate different frequencies of a song's latent spectrum with musical attributes like genre, chords, tempo, and pitch.} We select one song at a time to analyze. An input song is chosen from our validation split of GTZAN \citep{tzanetakis2002musical}. We generate hundreds of variations of the input song, while conditioning on different parts of its latent spectrum. We do this by performing a linear sweep over the latent frequency axis, conditioning on every 10-bin range of the latent DFT spectrum. We measure each generation's adherence to the input song along several axes, to determine how the latent frequency that we condition on affects which attributes are preserved.

First, we classify the genre of the generated variations. We train a classifier on our training split of GTZAN, which is a linear probe on VGGish embeddings \citep{hershey2017cnn}, and obtains 81.8\% accuracy on the validation set. Then, we apply our classifier to the generated variations, determining if the classifier's prediction of the generated variation matches the ground truth genre of the input song. For each frequency bin listed along the x-axis of our plot, we compute the accuracy across every variation whose condition included that bin. We normalize the curve to be between 0 and 1, so that it can be plotted alongside the other curves. 

Second, we measure the Tonnetz correlation between the variations and the reference. This provides a proxy to measuring changes in chords, since tonal centroid features are used to identify and compute the distance between chords \citep{milne2016empirically}. Again, we plot the Tonnetz correlation between input and variation against which frequency bins we condition on. We normalize this curve to be between 0 and 1 for the sake of plotting.

Third, we measure the pitch error using the Essentia \citep{bogdanov2013essentia} package. First, we use Essentia's algorithm to predict the predominant pitch (the pitch of the melody) in a song. Then, we compute the ``overall accuracy" metric described in \cite{salamon2014melody}, to measure how well the pitches of the generated variation match with the reference. Again, we plot the pitch accuracy against the latent frequencies that we condition on. We flip this curve vertically, so that `up' corresponds to higher preservation instead of higher error, and normalize the curve to have minimum 0 and maximum one.

Fourth, we estimate the BPM of the variations and the reference using Librosa \citep{mcfee2015librosa}. We compute the absolute tempo error between the reference and variants, again orienting the curve so that `up' corresponds to higher preservation, and normalizing the curve to be between 0 and 1.

We achieve the plots by applying Gaussian smoothing to all four curves. Note that unlike the blending and conditional generation experiments, we measure characteristics of the \emph{entire} generation versus the \emph{entire} reference, instead of bandpassing descriptor signals.

\clearpage
\section{Additional Experiments}

\subsection{\Rebuttaltwo{Ablations}}
\label{app:ablations}
\begin{figure}
    \centering
    \includegraphics[width=\linewidth]{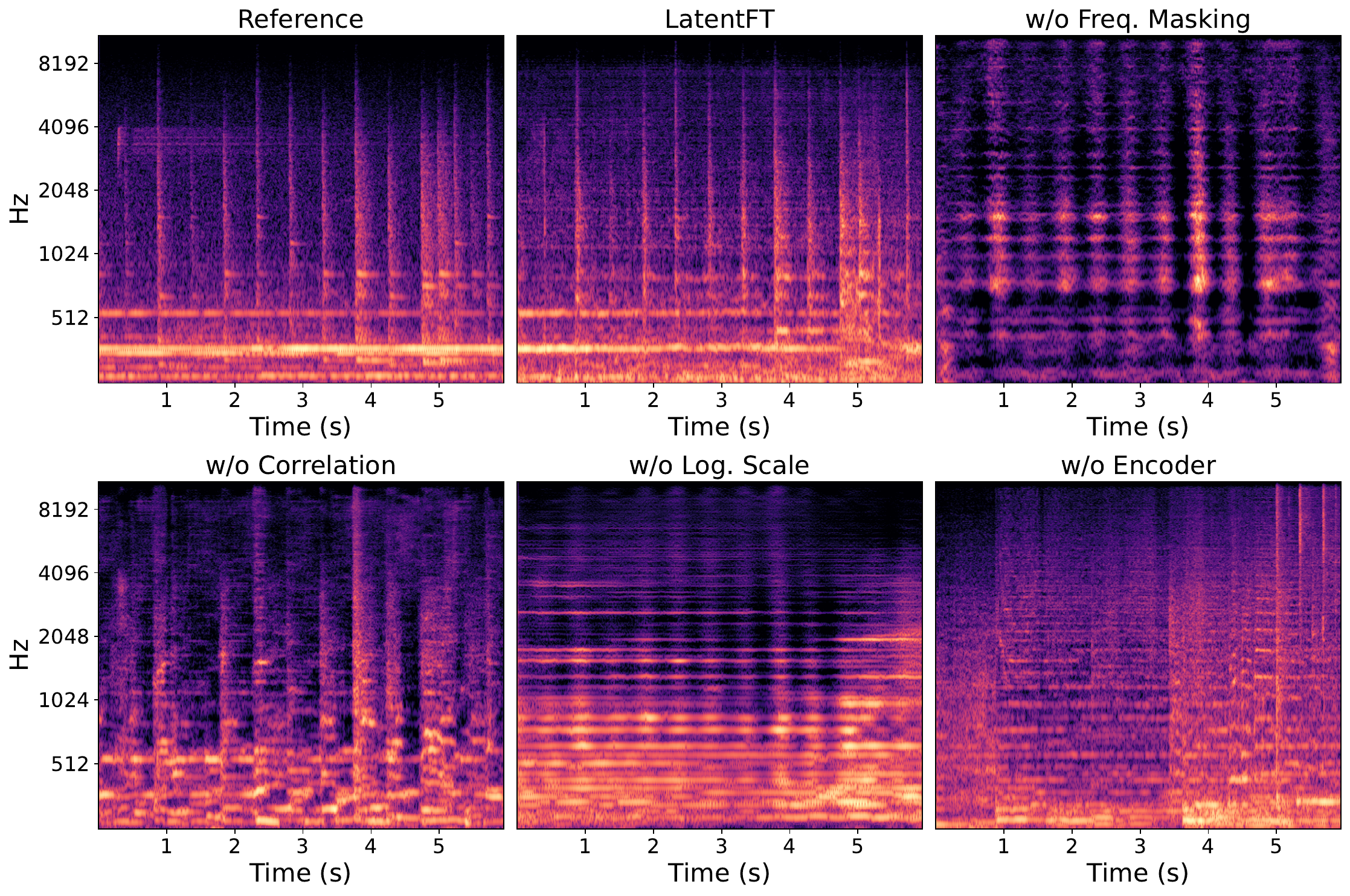}
    \caption{\Rebuttaltwo{A conditional generation example, where we take 0.68--2.70 Hz from the latent spectrum of the reference (top left). \name generates a variation capturing the rhythmic pattern near 2 Hz. The frequency-masking, correlation, and log-scaling ablations also have a pattern near 2 Hz, but the audio quality is much worse. The encoder ablation does not follow the conditioning.}}
    \label{fig:abl_cond_spec}
\end{figure}

\begin{figure}
    \centering
    \includegraphics[width=\linewidth]{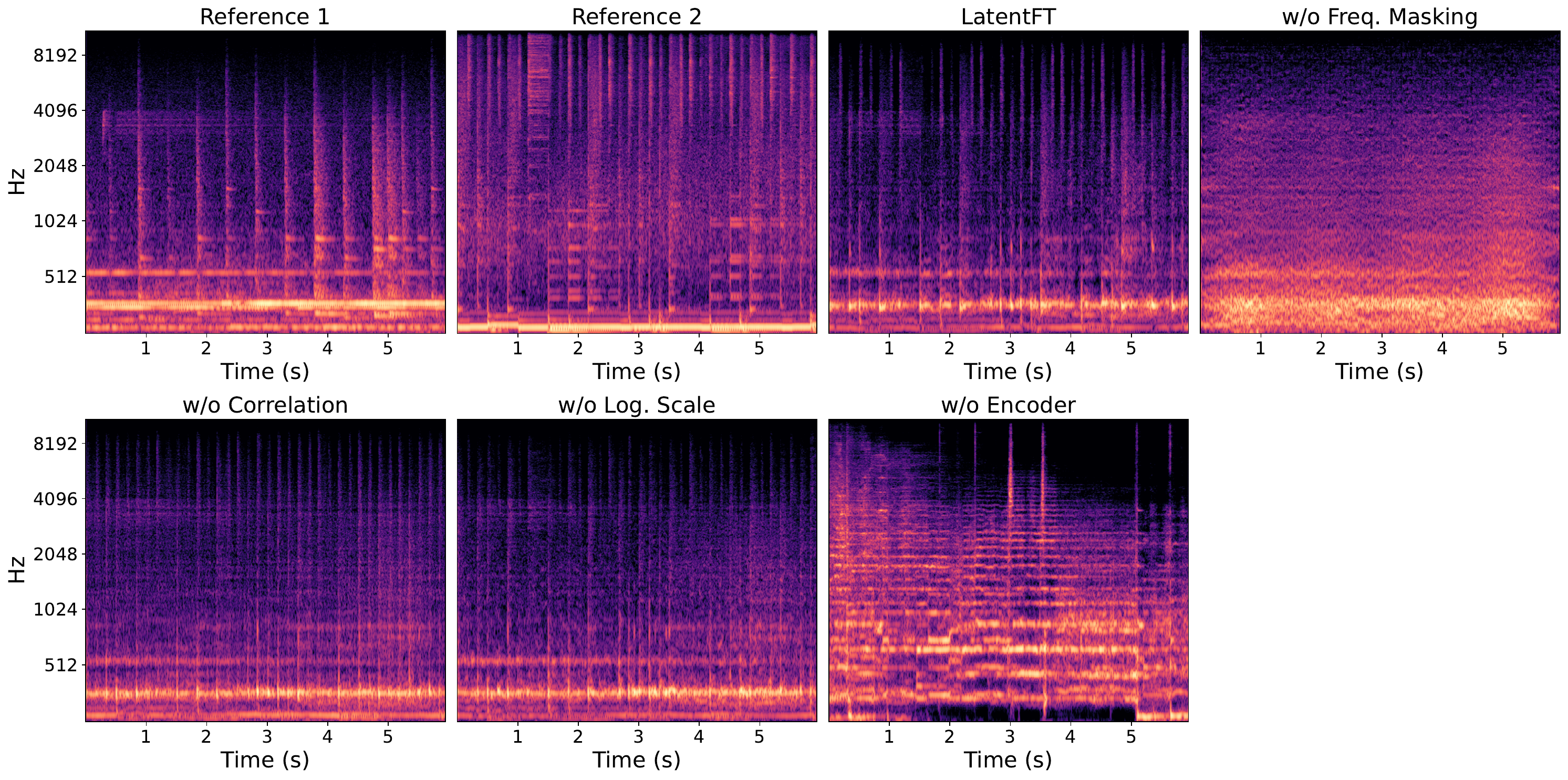}
    \caption{\Rebuttaltwo{A blending example, where we take 0--0.68 Hz from the first reference, and 10.78--43 Hz from the second reference. \name generates a variation that contains characteristics from both examples. For instance, the rapid rhythmic patterns of Reference 2 are retained, as well as the horizontal line from Reference 1. The correlation and log-scaling ablations retain some of these characteristics, while the encoder and frequency masking ablations ignore the references.}}
    \label{fig:abl_blend_spec}
\end{figure}

We ablate several components from \name-MLP \Rebuttaltwo{to demonstrate the necessity of each component. In these experiments, we train \name-MLP and each of its variants for 350k iterations, skipping the annealing phase. Quantitative results for the conditional generation task are shown in Table~\ref{tab:ablation-cond}. Quantitative results for the blending task are shown in Table~\ref{tab:ablation}. We also show example spectrograms for conditional generation in Fig.~\ref{fig:abl_cond_spec}, and example spectrograms for blending in Fig.~\ref{fig:abl_blend_spec}.}


\begin{table}[h]
\centering
\begin{tabular}
{>{\color{black}}l >{\color{black}}c>{\color{black}}c>{\color{black}}c>{\color{black}}c >{\color{black}}c}
    \toprule
    & \multicolumn{4}{c}{\Rebuttaltwo{\textbf{Adherence}}}
    & \multicolumn{1}{c}{\Rebuttaltwo{\textbf{Quality}}} \\

    \cmidrule(lr){2-5}
    \cmidrule(lr){6-6}

    & Loudness $\uparrow$ & Rhythm $\uparrow$ & Timbre $\downarrow$ & Harmony $\downarrow$ & FAD $\downarrow$ \\
            
    \midrule
    \name-MLP & 0.800 & \textbf{0.961} & \textbf{0.397} & \textbf{0.081} & \textbf{0.349} \\
    \:\:\:w/o Freq. Masking & 0.476 & 0.907 & 2.675 & 0.121 & 5.341 \\
    \:\:\:w/o Correlation & 0.694 & 0.932 & 1.284 & 0.109 & 2.744 \\
    \:\:\:w/o Log. Scale & 0.512 & 0.838 & 1.322 & 0.097 & 1.196 \\
    \:\:\:w/o Encoder & 0.028 & 0.565 & 3.569 & 0.130 & 0.846 \\
    \Rebuttalfour{\:\:\:w/ Bandpass Augmentation} & \Rebuttalfour{\textbf{0.861}} & \Rebuttalfour{0.953} & \Rebuttalfour{0.562} & \Rebuttalfour{0.084} & \Rebuttalfour{1.511}\\
    \bottomrule
    
\end{tabular}
\caption{\Rebuttaltwo{Ablation results on the \textbf{Conditional Generation} Task. Mel-Cepstral Distortion (Timbre) is divided by 100. Ablating any component of the model generally leads to worse audio quality and adherence.}}
\label{tab:ablation-cond}
\end{table}

\begin{table}[h]
\centering
\begin{tabular}{l cccc c}
    \toprule
    & \multicolumn{4}{c}{\textbf{Adherence}}
    & \multicolumn{1}{c}{\textbf{Quality}} \\

    \cmidrule(lr){2-5}
    \cmidrule(lr){6-6}

    & Loudness $\uparrow$ & Rhythm $\uparrow$ & Timbre $\downarrow$ & Harmony $\downarrow$ & FAD $\downarrow$ \\
            
    \midrule
    
    \name-MLP & \textbf{0.678} & 0.875 & \textbf{1.030} & \textbf{0.109} & 1.371 \\
    \:\:\:w/o Freq. Masking & 0.597 & \textbf{0.902} & 1.152 & 0.127 & 4.789 \\
    \:\:\:w/o Correlation & 0.635 & 0.885 & 1.167 & 0.115 & 2.534 \\
    \:\:\:w/o Log. Scale & 0.535 & 0.827 & 1.382 & 0.111 & 2.119 \\
    \:\:\:w/o Encoder & 0.030 & 0.539 & 4.026 & 0.147 & \textbf{0.854} \\
 \Rebuttalfour{\:\:\:w/ Bandpass Augmentation} & \Rebuttalfour{0.664} & \Rebuttalfour{0.885} & \Rebuttalfour{1.636} & \Rebuttalfour{0.117} & \Rebuttalfour{2.586} \\
    \bottomrule
    
\end{tabular}
\caption{\Rebuttaltwo{Ablation results on the \textbf{Blending} Task}. Mel-Cepstral Distortion (Timbre) is divided by 100. Ablating any component of the model generally leads to either significantly worse audio quality, or significantly worse adherence.}
\label{tab:ablation}
\end{table}

\myparagraph{\Rebuttaltwo{\textbf{Ablating Frequency Masking During Training.}}} First, we ablate frequency masking during training, applying only the inference-time user-specified mask. Previous methods apply frequency-masking post-hoc, to \emph{analyze} a pretrained model's latent space \citep{tamkin2020language}. \Rebuttaltwo{In Tables~\ref{tab:ablation-cond} and~\ref{tab:ablation} (``w/o Freq. Masking''), we see that removing frequency-masking during training results in a substantial degradation in audio quality. Without masking during training, the decoder does not learn how to reconstruct music from frequency-masked latents, and fails to generate high-quality audio from frequency-masked latents during inference.} This ablation also shows post-hoc masking is insufficient for coherent audio \emph{synthesis}, which requires incorporating masking during training. 

\begin{figure}[h]
    \centering
    \begin{minipage}[t]{0.45 \linewidth}
        \centering
        \includegraphics[width=\linewidth]{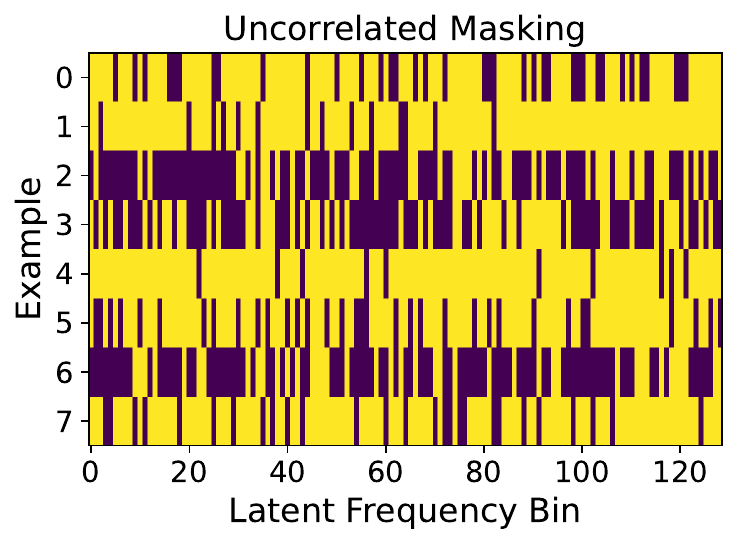}
        \caption{\Rebuttaltwo{Example masks where there is no correlation between the scores associated with each frequency bin. The masks are speckled and erratic.}}
        \label{fig:uncorrmask}
    \end{minipage}
    \hfill
    \centering
    \begin{minipage}[t]{0.45 \linewidth}
        \centering
        \includegraphics[width=\linewidth]{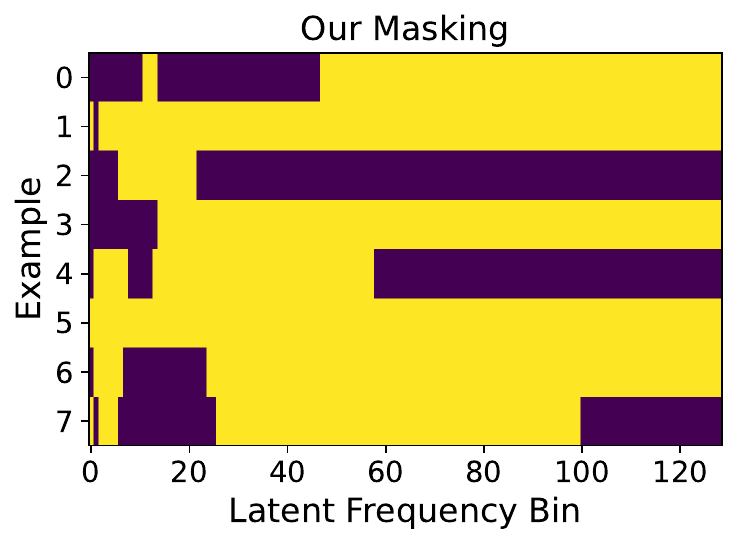}
        \caption{\Rebuttaltwo{Example masks from our masking strategy, where bin scores are locally correlated after being mapped to a logarithmic axis. The mask forms contiguous regions.}}
        \label{fig:ourmask}
    \end{minipage}
\end{figure}

\myparagraph{\Rebuttaltwo{\textbf{Ablating Correlations between Bins.}}} Next, we ablate correlations between frequency bins. \Rebuttaltwo{As explained in Sec.~\ref{subsec:masking}, we use locally correlated scores to mask frequency bins. If we mask each bin independently, we will end up with speckled, erratic masks, where unmasked bins and masked bins are next to each other. This is shown in Fig.~\ref{fig:uncorrmask}. Unmasked bins provide strong local cues to nearby masked bins, making the reconstruction/denoising task easier. In contrast, masks generated from locally correlated scores are shown in Fig.~\ref{fig:ourmask}. Our strategy of correlating scores results in large, contiguous regions of unmasked and masked bins, which makes the learning task more difficult, and better reflects inference-time, user-specified masks. Tables ~\ref{tab:ablation-cond} and~\ref{tab:ablation} (``w/o Correlation") verify that using an uncorrelated mask results in substantial degradations to audio quality.
}

\myparagraph{\Rebuttaltwo{\textbf{Ablating Logarithmic Scaling of Latent Frequency Axis.}}} We also ablate the logarithmic scaling of the frequency axis, discussed in Sec.~\ref{subsec:masking}. \Rebuttaltwo{Our intuition for using a logaritmic scaling is as follows: Most structured signals have a $1/f$-spectrum, meaning that the energy at high frequencies is much lower than the energy at low frequencies. Thus, a “group” of low-frequency bins will contain much more energy than a “group” of high-frequency bins of equal width. To counterbalance this effect, we encourage high-frequency “groups” to be wider, by mapping the frequency bins to a logarithmic scale before computing correlations between bins. This reflects the fact that $1/f$-spectra have equal energy per-octave. Indeed, removing logarithmic scaling reduces both quality and adherence in both the conditional generation and blending tasks, shown in Tables ~\ref{tab:ablation-cond} and~\ref{tab:ablation} (``w/o Log. Scale").}

\myparagraph{\Rebuttaltwo{\textbf{Ablating Encoder.}}} We also ablate the encoder, applying frequency-masking to the audio waveform directly, to show that our model's representations capture things the waveform cannot. Ablating the encoder results in poor adherence, but better audio quality \Rebuttaltwo{in the case of blending (Tables ~\ref{tab:ablation-cond} and~\ref{tab:ablation} (``w/o Encoder")}. Note that allowing for poor adherence improves audio quality, since the generation is less constrained. \Rebuttaltwo{Since there is very little information in the waveform for frequencies we condition on (0--43 Hz), removing the encoder is almost like running an unconditional model.}

\myparagraph{\Rebuttalfour{Random Bandpass Augmentation.}}
\Rebuttalfour{Lastly, we would like to test the necessity of using a DFT-based mask as our latent augmentation, instead of another frequency-aware latent-space augmentation. Thus, instead of applying a DFT Mask to the latent space during training and inference, we apply a randomized bandpass filter to the latent space during training, and a user-specified one during inference. We found that this resulted in some training instability, requiring several restarts. We believe the orthogonality of the DFT is helpful for training stability: From a theoretical perspective, DFT-masking in the forwards pass results in applying the same DFT mask to the upstream gradient in the backwards pass. In the backwards pass, the DFT mask can thus be interpreted as masking out orthogonal components of the upstream gradient, while leaving the unmasked components of the full-band gradient intact.
}
\myparagraph{\Rebuttaltwo{\textbf{Example Spectrograms.}}}

\Rebuttaltwo{
Fig.~\ref{fig:abl_cond_spec} shows example spectrograms for the ablations on conditional generation, and Fig.~\ref{fig:abl_blend_spec} shows example spectrograms for blending. The figures show that many of the baselines fail to generate coherent audio. The ablation without the encoder generates coherent audio, but fails to follow the condition(s). Please refer to the figure captions for more details.}


\clearpage
\subsection{\Rebuttaltwo{Results on More Datasets}}
\label{app:more_datasets}
\Rebuttaltwo{
To demonstrate generality, we also use \name to perform conditional generation and blending on two other datasets: GTZAN and Maestro. The GTZAN dataset was previously used for the interpretability experiment in Sec.~\ref{sec:interpretability}, and is described in Appendix~\ref{app:datasets}. The Maestro dataset \citep{hawthorne2018enabling} is
a large collection of over 200 hours of aligned piano performance audio and MIDI from the International Piano-e-Competition.
}

\Rebuttaltwo{
Taking our \name-MLP model trained on the MTG-Jamendo training set, we evaluate the model on 1024 5.9-second clips from both the GTZAN and Maestro datasets. The results for GTZAN are show in Table~\ref{tab:gtzan-condandblend}, and the results for Maestro are shown in Table~\ref{tab:maestro-condandblend}. Although \name performs worse in terms of audio quality compared to our evaluations on MTG-Jamendo, we find that it outperforms our baselines on both GTZAN and Maestro. This indicates that \name can work on recordings that are only piano, or on datasets with a diverse set of genres.
}
\begin{table*}[h]
\centering
\resizebox{\linewidth}{!}{%
\begin{tabular}{>{\color{black}}l *{10}{>{\color{black}}c}}
    \toprule
    %
    & \multicolumn{5}{c}{\Rebuttaltwo{\textbf{Conditional Generation}}}
    & \multicolumn{5}{c}{\Rebuttaltwo{\textbf{Blending}}} \\
    \cmidrule(lr){2-6}
    \cmidrule(lr){7-11}

    & \multicolumn{4}{c}{\Rebuttaltwo{Adherence}}
    & \multicolumn{1}{c}{\Rebuttaltwo{Quality}}
    & \multicolumn{4}{c}{\Rebuttaltwo{Adherence to Both Inputs}}
    & \multicolumn{1}{c}{\Rebuttaltwo{Quality}} \\        
    \cmidrule(lr){2-5}
    \cmidrule(lr){6-6}
    \cmidrule(lr){7-10}
    \cmidrule(lr){11-11}

    & Loud. $\uparrow$ & Rhyth. $\uparrow$ & Timb. $\downarrow$ & Harm. $\downarrow$ & FAD $\downarrow$ & Loud. $\uparrow$ & Rhyth. $\uparrow$ & Timb. $\downarrow$ & Harm. $\downarrow$ & FAD $\downarrow$\\
    
    \midrule
    Vampnet & - & - & - & - & 5.748 & - & - & - & - & 7.173 \\
Guidance Gradients & 0.585 & 0.825 & 1.470 & 0.094 & 1.368 & 0.611 & 0.850 & 1.643 & 0.105 & 1.961 \\
ILVR & 0.628 & 0.852 & 0.730 & 0.088 & 1.873 & 0.672 & 0.877 & \textbf{0.744} & 0.097 & 3.137 \\
DAC & 0.723 & 0.845 & 4.045 & 0.191 & 8.810 & 0.610 & 0.794 & 4.115 & 0.212 & 7.162 \\
Spectrogram & 0.503 & 0.876 & 1.873 & 0.128 & 8.734 & 0.402 & 0.840 & 2.972 & 0.111 & 8.397 \\
Cross Synthesis & - & - & - & - & - & - & - & - & - & 2.884 \\
\midrule
\name-MLP & 0.840 & 0.965 & \textbf{0.356} & \textbf{0.073} & \textbf{0.844} & \textbf{0.721} & 0.885 & 0.970 & \textbf{0.095} & 1.987 \\
\name-UNet & \textbf{0.855} & \textbf{0.967} & 0.377 & \textbf{0.073} & 0.905 & 0.714 & \textbf{0.891} & 1.056 & \textbf{0.095} & \textbf{1.926} \\
\bottomrule
\end{tabular}
}%
\caption{\Rebuttaltwo{Results on Conditional Generation and Blending on the GTZAN dataset. Compared to baselines, \name achieves superior adherence and audio quality, demonstrating the generality of \name when it comes to new datasets with multiple genres. Mel-Cepstral Distortion (Timbre) is divided by 100. 
 The Masked Token Model and Cross Synthesis baselines do not offer frequency-based controls, so we do not compute adherence. Cross Synthesis also only applies to the blending task.}}
\label{tab:gtzan-condandblend}
\end{table*}

\begin{table*}[h]
\centering
\resizebox{\linewidth}{!}{%
\begin{tabular}{>{\color{black}}l *{10}{>{\color{black}}c}}
\toprule
    %
    & \multicolumn{5}{c}{\Rebuttaltwo{\textbf{Conditional Generation}}}
    & \multicolumn{5}{c}{\Rebuttaltwo{\textbf{Blending}}} \\
    \cmidrule(lr){2-6}
    \cmidrule(lr){7-11}

    & \multicolumn{4}{c}{\Rebuttaltwo{Adherence}}
    & \multicolumn{1}{c}{\Rebuttaltwo{Quality}}
    & \multicolumn{4}{c}{\Rebuttaltwo{Adherence to Both Inputs}}
    & \multicolumn{1}{c}{\Rebuttaltwo{Quality}} \\        
        
    \cmidrule(lr){2-5}
    \cmidrule(lr){6-6}
    \cmidrule(lr){7-10}
    \cmidrule(lr){11-11}

    & Loud. $\uparrow$ & Rhyth. $\uparrow$ & Timb. $\downarrow$ & Harm. $\downarrow$ & FAD $\downarrow$ & Loud. $\uparrow$ & Rhyth. $\uparrow$ & Timb. $\downarrow$ & Harm. $\downarrow$ & FAD $\downarrow$\\
    
    \midrule
    Vampnet & - & - & - & - & 11.914 & - & - & - & - & 14.887 \\
    Guidance Gradients & 0.530 & 0.795 & 1.483 & 0.116 & 8.588 & 0.557 & 0.824 & 1.606 & 0.133 & 6.221 \\
    ILVR & 0.580 & 0.817 & 0.976 & 0.118 & 9.923 & 0.627 & 0.857 & 1.007 & 0.131 & 10.018 \\
    DAC & 0.729 & 0.835 & 4.088 & 0.243 & 11.745 & 0.639 & 0.776 & 3.720 & 0.297 & 11.614 \\
    Spectrogram & 0.413 & 0.853 & 1.981 & 0.152 & 14.208 & 0.330 & 0.817 & 2.640 & 0.157 & 14.131 \\
    Cross Synthesis & - & - & - & - & - & - & - & - & - & 3.139 \\
    \midrule
    \name-MLP & 0.809 & 0.967 & \textbf{0.553} & \textbf{0.085} & \textbf{0.667} & 0.689 & 0.892 & \textbf{0.886} & \textbf{0.121} & 2.767 \\
    \name-UNet & \textbf{0.830} & \textbf{0.968} & 0.590 & \textbf{0.085} & 0.865 & \textbf{0.710} & \textbf{0.899} & 0.943 & 0.124 & \textbf{2.708} \\
    \bottomrule
\end{tabular}
}%
\caption{\Rebuttaltwo{Results on Conditional Generation and Blending on the Maestro dataset. Even though the Maestro dataset is only piano recordings, \name demonstrates super audio quality and adherence compared to baselines. Mel-Cepstral Distortion (Timbre) is divided by 100.
The Masked Token Model and Cross Synthesis baselines do not offer frequency-based controls, so we do not compute adherence. Cross Synthesis also only applies to the blending task.}}
\label{tab:maestro-condandblend}
\end{table*}

\clearpage
\subsection{\Rebuttaltwo{More Interpretability Results}}
\label{app:more-interpretability}

\begin{figure}[h]
    \centering
    \includegraphics[width=\linewidth]{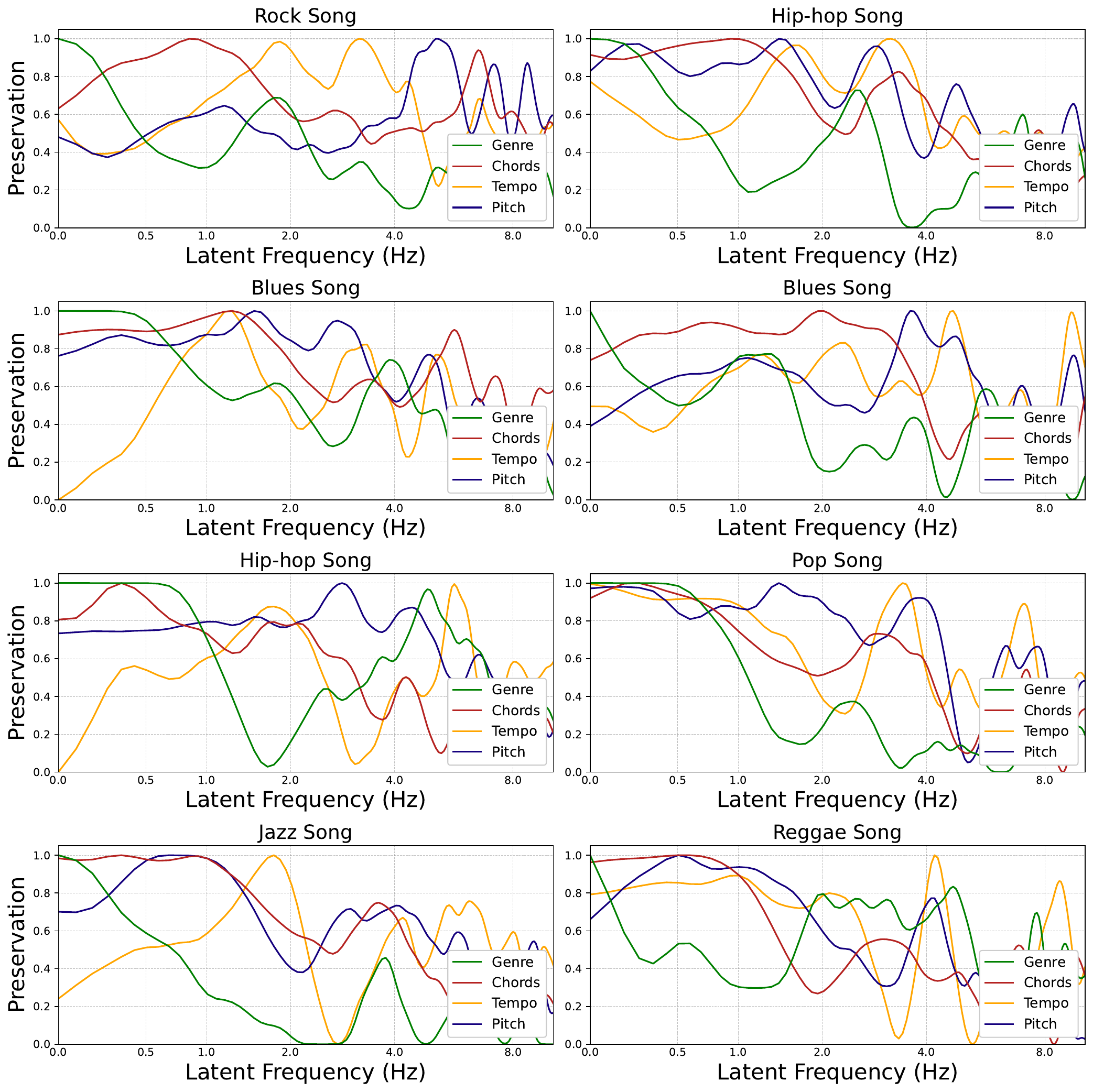}
    \caption{\Rebuttaltwo{More Sweep Examples. Songs are taken from the GTZAN dataset. Generally, genre tends to be a global characteristic, lying around 0 Hz. Chord changes also lie in the low end of the latent spectrum, while tempo and pitch are associated with higher latent  frequencies. Please refer to Sec.~\ref{sec:interpretability} for our motivations behind this experiment, and Appendix~\ref{app:sweep} for implementation details.}}
    \label{fig:more_sweep_examples}
\end{figure}

\Rebuttaltwo{Our interpretability experiment, introduced in Sec.~\ref{sec:interpretability}, attributes parts of a particular song's latent spectrum with musical characteristics like genre, chords, tempo, and pitch. In this section, we present more examples where we analyze individual songs, and plot how well conditioning on various latent frequencies in the song preserve genre, chords, tempo, and pitch. These extra plots are show in Fig.~\ref{fig:more_sweep_examples}. Across several musical styles, we see the trend that genre tends to lie in the frequency range around 0 Hz, indicating that it is a global characteristic. Chord changes also occur at low frequencies, with peak preservation between 0.25--2 Hz. Tempo and pitch occur at higher latent frequencies, since prominent rhythmic and melodic patterns are typically more rapid than chord changes. Please refer to Appendix~\ref{app:sweep} for implementation details.
}

\clearpage
\subsection{\Rebuttaltwo{Removing the Latent DFT}}
\begin{figure}[h]
    \centering
    \includegraphics[width=0.7\linewidth]{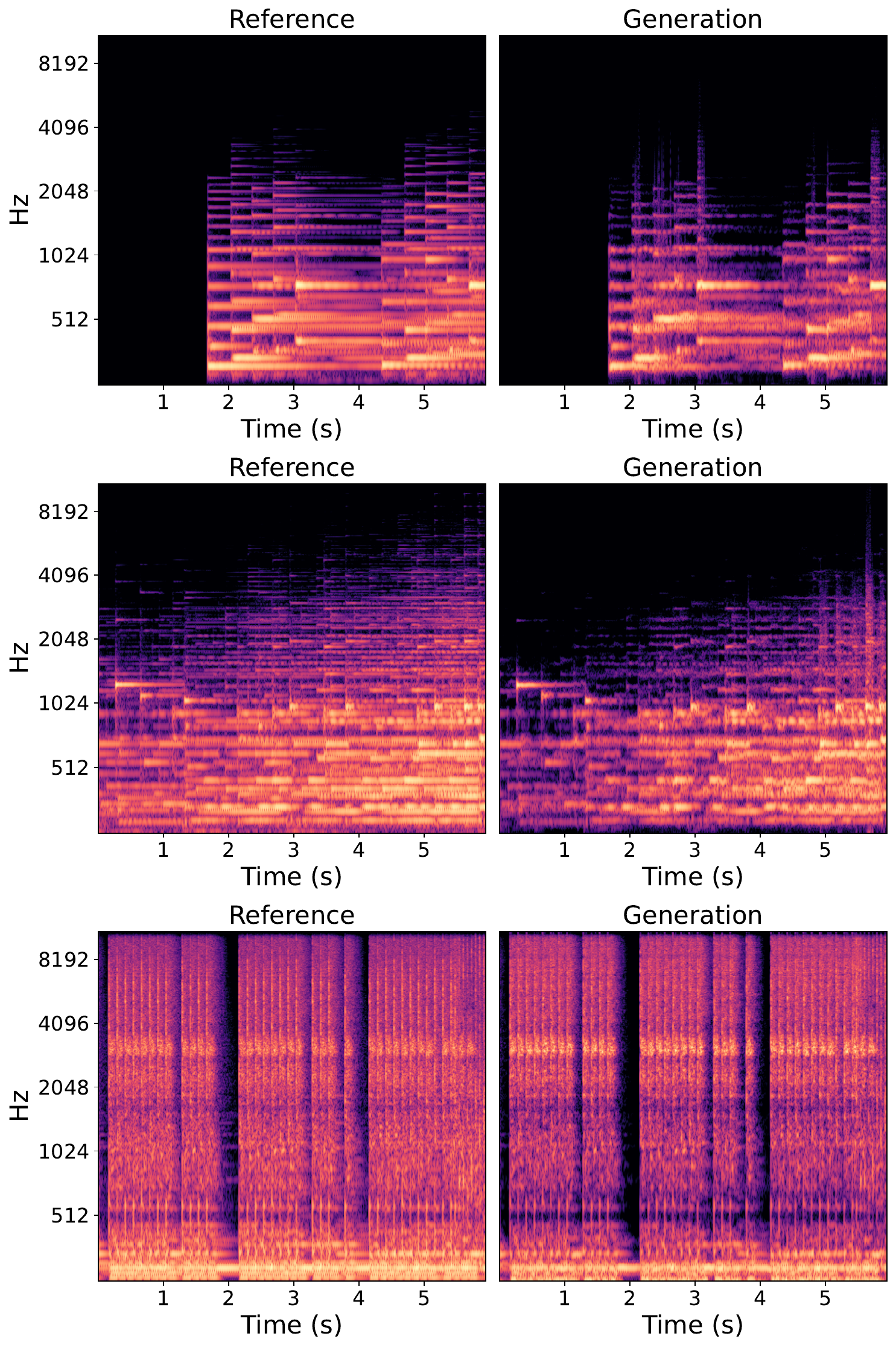}
    \caption{\Rebuttaltwo{Mel-spectrograms where we remove the DFT during both training and inference. During inference, we condition the diffusion process on the full latent sequence $\vz$ derived from a reference (left). This reconstructs the input without creating a variation (right).}}
    \label{fig:removing-latent-dft}
\end{figure}

\Rebuttaltwo{
In this experiment, we remove the Latent DFT entirely from both training and inference. During training, the model tries to reconstruct the input from the full latent sequence $\vz$. During inference, the full latent sequence $\vz$ remains unmasked. This is similar to the original Diffusion Autoencoder from \cite{preechakul2022diffusion}. We find that without frequency masking, the decoder reconstructs in the input without generating any interesting variations, as show in Fig.~\ref{fig:removing-latent-dft}. For audio examples, refer to the \href{https://masonlwang.com/latentfouriertransform/}{website} under ``Removing DFT Masking''.}

\clearpage
\subsection{Per-Band Error}
\begin{figure}[H]
    \centering
    \includegraphics[width=0.7\linewidth]{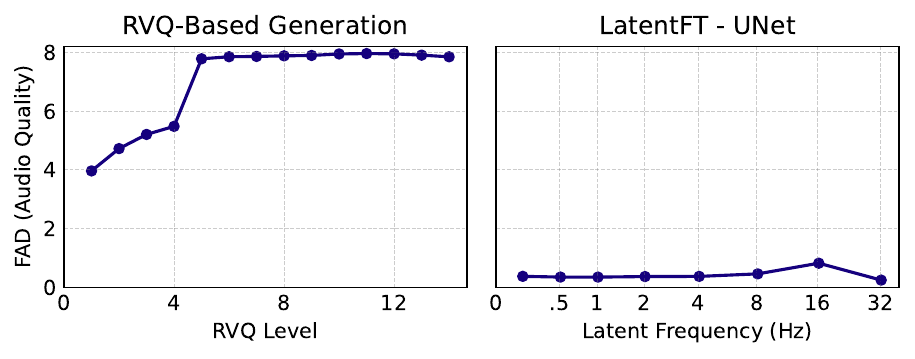}
    \caption{Conditioning on various RVQ layers in the Vampnet Model (left) and on various latent frequencies in our model (right). Our model maintains generation quality even when conditioning on finer-scale features.}
    \label{fig:per-band-error}
\end{figure}

We show in Fig.~\ref{fig:per-band-error} that conditioning on mid-scale or fine-scale RVQ levels leads to a rapid degradation in audio quality. On the left, we generate audio using the Masked Token Model baseline \citep{garcia2023vampnet}, which contains 14 RVQ levels in total. We condition on each of the levels individually, and observe a degradation in quality as we condition on finer and finer tokens. On the right, we show a comparison with our model, conditioning on different latent frequency bands instead of different RVQ layers. As we condition on higher and higher frequencies (smaller timescales), the audio quality does not degrade. The metrics shown are averaged across 1024 songs from the MTG-Jamendo test set.

\section{\Rebuttaleight{Additional Related Work}}
\myparagraph{Separating Info by Scale.} Our work relates to learned multiscale representations that attempt to separate information by scale. Hierarchical VAEs attempt to model the data distribution using a (often multiscale) stack of latent variables. However, \cite{zhao2017learning} show theoretically and experimentally that most hierarchical VAEs \citep{sonderby2016ladder, gulrajani2016pixelvae, bachman2016architecture} have difficulty separating information between levels. They propose an alternative multi-network architecture under the assumption that deeper networks encode more abstract features, while shallower ones will encode simpler ones. Using these networks, they show they can vary features of an image across a few (e.g. 4) scales independently. Still, the exact scale that each network corresponds to depends on the data distribution. We extend this by 1) providing a \emph{continuous} scale axis and 2) providing an \emph{intuitive, non-heuristic} way of specifying scales via Hz.

\myparagraph{Generative Audio Equalizer.} Similar to our work, \cite{moliner2024diffusion} introduce a diffusion-based generative audio equalizer. While this work generates content at selected \emph{audible} frequencies, we generate content at \emph{latent} frequencies.

\myparagraph{Other Uses of the Fourier Transform in Deep Learning.} The Fourier transform has also been used in CNNs to accelerate convolutions \citep{mathieu2013fast, ding2017circnn}. Audio signals are also ubiquitously represented in the frequency domain, as are MRI images \citep{passigan2024analyzing}.

\Rebuttaleight{\myparagraph{AudioMAE.} Another work in audio that uses a masking strategy during training is AudioMAE \citep{huang2022masked}. This work builds off of masked autoencoders for images \citep{he2022masked}. Here, a neural network tries to reconstruct an audio spectrogram after many time-frequency patches have been masked. This task allows the network to learn representations that are useful for classification, event detection, and retrieval. While AudioMAE masks random time-frequency bins, \name masks random bins in the latent spectrum.
}

\clearpage
\section{Extended Background}
\subsection{DFT}
\label{app:dft}
Here, we derive Eq.~\ref{eq:realidft}:
\begin{align*}
    \vx &= \frac{1}{N} \sum_{k=0}^{N-1} \bm{X}[k] \bm{w}_{k} \\
    \vx &= \frac{1}{N} \sum_{k=0}^{N-1} \bm{X}[k] e^{j(2\pi k / N) n}
\end{align*}
We consider the case where $N$ is odd. By the periodicity of complex sinusoids:
$$
    \vx = \frac{1}{N} \sum_{k=-\floor{N/2}}^{\floor{N/2}} \bm{X}[k] e^{j(2\pi k / N) n} \nonumber \\
$$
We can combine the $k < 0$ terms with the $k > 0$ terms in the sum like so:
\begin{align*}
    \vx &= \frac{1}{N} \sum_{k=1}^{\floor{N/2}}  \left[\bm{X}[k] e^{j(2\pi k / N) n} + \bm{X}[-k] e^{-j(2\pi k / N) n} \right] + \frac{\bm{X}[0]}{N}  \nonumber \\
\end{align*}
Where we take the $k=0$ out of the sum. The DFT of a real-valued signal is Hermitian, meaning that $\bm{X}[-k] = \bm{X}^*[k]$: 
\begin{align*}
    \vx &= \frac{1}{N} \sum_{k=1}^{\floor{N/2}}  \left[\bm{X}[k] e^{j(2\pi k / N) n} + \bm{X}^*[k] e^{-j(2\pi k / N) n} \right] + \frac{\bm{X}[0]}{N}  \nonumber \\
\end{align*}
As its complex conjugate, $\bm{X}^*[k]$ has the same magnitude as $\bm{X}[k]$, but the opposite phase. We can split $\bm{X}[k]$ into its magnitude $|\bm{X}[k]|$ and phase $\phi_k$, and do the same for $\bm{X}^*[k]$:
$$
    \bm{X}[k] = |\bm{X}[k]| e^{j\phi_k}
$$
$$
    \bm{X}^*[k] = |\bm{X}[k]| e^{-j\phi_k}
$$
Plugging these into our formula:
\begin{align*}
    \vx &= \frac{1}{N} \sum_{k=1}^{\floor{N/2}}  \left[ |\bm{X}[k]| e^{j\phi_k} e^{j(2\pi k / N) n} + |\bm{X}[k]| e^{-j\phi_k} e^{-j(2\pi k / N) n } \right] + \frac{\bm{X}[0]}{N}  \nonumber \\
    \vx &= \frac{1}{N} \sum_{k=1}^{\floor{N/2}} |\bm{X}[k]| \left[ e^{j\left((2\pi k / N) n + \phi_k\right)} + e^{-j\left((2\pi k / N) n + \phi_k\right)} \right] + \frac{\bm{X}[0]}{N}  \nonumber \\
\end{align*}
Using Euler's Formula:
\begin{align*}
    \vx[n] &=  \frac{1}{N} \sum_{k=1}^{\floor{N/2}} 2 | \bm{X}[k] |  \cos \left( \frac{2 \pi}{N} k n + \phi_k \right) + \frac{\bm{X}[0]}{N} \nonumber \\
\end{align*}
This can be expressed as:
\begin{align*}
\vx[n] &=  \sum_{k=0}^{\floor{N/2}} A_k \cos \left( 2 \pi\frac{k}{N} n + \phi_k \right) \nonumber \\
\end{align*}
Which is the form that we desire. Observe that the constant term outside of the sum re-enters the sum as a constant cosine ($k=0$). The case where $N$ is even is quite similar, but includes another term (the Nyquist term), which is always a real cosine.

\section{LLM Usage}
We used LLMs to help us improve the writing of our paper, for instance, by finding synonyms for certain words or for finding more concise ways to phrase particular ideas. We also used LLMs as a search tool to help us find related work, but relied on our own interpretation of the work after references were provided.

\end{document}